\documentclass[aps,prc,reprint,superscriptaddress,floatfix,showpacs,nofootinbib]{revtex4-1}

\usepackage{epsfig}
\usepackage{dcolumn}
\usepackage{bm}
\usepackage{amssymb}
\usepackage{amsmath}

\usepackage{mathtools}
\usepackage{mathrsfs}

\usepackage{amsfonts}
\usepackage{soul}
\usepackage[usenames,dvipsnames]{color}

\begin{document}

\title{Impact of Multiplicity Fluctuations on Entropy Scaling Across System Size}

\author{P. Carzon}
\email[Email: ]{pcarzon2@illinois.edu}
\affiliation{Illinois Center for Advanced Studies of the Universe \&  Department of Physics, 
University of Illinois at Urbana-Champaign, Urbana, IL 61801, USA}
\author{M. Sievert}
\email[Email: ]{msievert@nmsu.edu}
\affiliation{Department of Physics, New Mexico State University, Las Cruces, NM 88003, USA}
\author{J. Noronha-Hostler}
\email[Email: ]{jnorhos@illinois.edu}
\affiliation{Illinois Center for Advanced Studies of the Universe \&  Department of Physics, 
University of Illinois at Urbana-Champaign, Urbana, IL 61801, USA}

\date{\today}
\begin{abstract}
The initial state is one of the greatest uncertainties in heavy-ion collisions.  A model-agnostic approach is taken in the phenomenological Trento framework which constrains parameters using Bayesian analysis. However, the color-glass condensate (CGC) effective theory predicts initial energy densities that lie outside the recent Bayesian analyses due, in part, to the assumption in Trento of event-by-event multiplicity fluctuations following a $\Gamma$ distribution. We compare the Trento-preferred $\sqrt{T_{A}T_{B}}$ scaling to CGC-like $T_{A}T_{B}$ scaling coupled with log-normal fluctuations in $AuAu$ and $dAu$ collisions and find there is a significant impact on the multiplicity distributions and on the eccentricities, which may affect the extraction of viscosity in small systems.
\end{abstract}

\maketitle

\section{Introduction}
It is well known that the anisotropic flow harmonics $v_n$ measured in heavy-ion collisions can be described by the deterministic response of viscous hydrodynamics to the initial geometry of the collision \cite{Noronha-Hostler:2015uye, Niemi:2015voa, Adam:2015ptt,Song:2010mg, Bozek:2012qs, Gardim:2012yp, Bozek:2013uha, Niemi:2015qia, Ryu:2015vwa, McDonald:2016vlt, Bernhard:2016tnd, Gardim:2016nrr, Alba:2017hhe, Giacalone:2017dud, Eskola:2017bup, Weller:2017tsr, Schenke:2019ruo}.  The initial-state geometry in an event can be quantified using the eccentricities, $\varepsilon_{n} = | \langle e^{i n \phi} \rangle |$, which are strongly correlated with the flow harmonics, $v_n$, through quasi-linear response \cite{Gardim:2011xv, Gardim:2014tya}.  The transport parameters of hydrodynamics, such as the shear viscosity to entropy density ratio $\eta/s$, determine the damping of the initial eccentricities into the final observed flow harmonics.  Consequently, different choices of initial conditions, matched to the same experimental flow data, will yield different extractions of the viscosity and other transport parameters \cite{Luzum:2009sb,Bernhard:2015hxa,Heinz:2011kt}.  Thus, the correct description of the initial state is crucial to an accurate extraction of the properties of the Quark-Gluon Plasma (QGP).

One approach is to implement theory-agnostic phenomenological models which capture the most important features of the initial-state geometry.  An early implementation of this philosophy was the two-component Glauber model \cite{Kharzeev:2000ph, Goldschmidt:2015kpa}, in which the initial-state geometry was constructed as a superposition of one term proportional to the distribution of wounded nucleons and another proportional to the ``binary collision density.''  While this model adequately described the multiplicity distribution and event geometry in round nuclei, it tended to significantly overpredict the modification of these distributions in central collisions of deformed nuclei such as uranium \cite{Voloshin:2010ut, Goldschmidt:2015kpa}.  In particular, the binary collision density contribution generally leads to a large  difference in the multiplicity of tip-on-tip collisions (scaling like $N_{part}^2$) compared to side-on-side collisions (scaling like $N_{part}$).  The resulting slope of the elliptic flow $v_2 \{2\}$ versus multiplicity in such a model was then much too steep in comparison with the experimental data measured at STAR \cite{Pandit:2013uiv, wang:2014qxa, adamczyk:2015obl}, leading to the exclusion of the two-component Glauber model as a viable description of the initial state.  

Inspired by these conclusions about the absence of strong multiplicity scaling with $N_{part}^2$, the authors of the phenomenological model Trento \cite{Moreland:2014oya} implemented a general scaling criterion enforcing that the multiplicity of tip-on-tip collisions scale comparably to that of side-on-side collisions:
\begin{align}   \label{e:genmean}
    s (T_A , T_B) \propto \left( \frac{T_A^p + T_B^p}{2} \right)^{1/p} ,
\end{align}
where $s$ is the initial entropy density produced by colliding nuclear profiles $T_A$ and $T_B$ and $p$ is any real number.  The generalized mean \eqref{e:genmean} assumed in Trento explicitly enforces a homogeneous scaling
\begin{align}   \label{e:trentoscaling}
    s (N \, T_A , N \, T_B) = N \: s (T_A , T_B)
\end{align}
which reduces the strong multiplicity differences between tip-on-tip and side-on-side collisions of deformed nuclei, reducing the slope of $v_2$ versus $N_{ch}$ and bringing the theory predictions more in line with the experimental data.  Interestingly, more recent work which modified Trento to include a type of binary collision term $s (T_A , T_B) \propto T_A T_B$ showed that an initial condition which explicitly violates the scaling \eqref{e:trentoscaling} is able to describe the ultracentral UU data after all \cite{Noronha-Hostler:2019ytn}. The linear entropy scaling used in this work \cite{Noronha-Hostler:2019ytn} is strikingly similar to the trend of CGC based simulations producing linear energy scaling \cite{Nagle:2018ybc, Lappi:2006hq, Chen:2015wia, Romatschke:2017ejr}. To draw a closer comparison to CGC, we treat the nuclear profiles from trento for $T_A T_B$ as proportional to energy and convert to entropy using a conformal equation of state. This is a simplification of what should be done, but it is sufficient to make comparisons between the two scalings. To be precise, we take the linear scaling to be $s (T_A , T_B) \propto (T_A T_B)^{3/4}$.  This may perhaps indicate that the flaw in the two-component Glauber model may have less to do with its multiplicity scaling than with the spatial profile with which the binary collision density was assumed to be deposited. In the meantime, other issues arose with the two-component Glauber model because it does not capture the fluctuations of $v_n$ correctly on an event-by-event basis \cite{Renk:2014jja,Giacalone:2017uqx}. 

A different approach to initial condition models is to perform a microscopic calculation in a particular theory to determine the initial energy or entropy density produced at early stages of the collision.  Of particular note are calculations based on the color-glass condensate (CGC) effective field theory for the high gluon densities realized in heavy-ion collisions \cite{McLerran:1993ni, McLerran:1993ka, McLerran:1994vd}.  These microscopic calculations introduce significant new sources of event-by-event fluctuations in the color degrees of freedom of the gluon fields, as implemented for instance in the successful IP-Glasma model \cite{Schenke:2012wb} based on classical Yang-Mills color fields.  However, while CGC models can successfully describe the initial geometry produced in heavy-ion collisions, the importance of the event-by-event color field fluctuations in that outcome is less clear.  Interestingly, it has been shown that at early times close to $\tau = 0$ after the collision, the initial energy density produced \textit{after averaging over these color fluctuations} is precisely proportional to a binary collision-type term $\epsilon \propto T_A T_B$ \cite{Nagle:2018ybc, Lappi:2006hq, Chen:2015wia, Romatschke:2017ejr}.  As such, this implies that these CGC-motivated initial conditions fall outside the scaling ansatz \eqref{e:trentoscaling} and are not captured within the flexible Trento framework, even though these CGC models have been able to successfully describe the multiplicity distributions produced in dilute-dilute \cite{McLerran:2015qxa}, dilute-dense \cite{Mace:2018vwq}, and dense-dense collisions \cite{Schenke:2012wb}.

Recently Bayesian analyses \cite{Moreland:2018gsh,Everett:2020xug,Nijs:2020roc} have used Trento \cite{Moreland:2014oya}, and demonstrated a preference for $p=0$ within the functional form \eqref{e:genmean}, corresponding to an initial entropy density $s \propto \sqrt{T_{A}T_{B}}$ where $T_A , T_B$ are the nuclear thickness functions   \cite{Moreland:2018gsh}. This Bayesian analysis also assumed a particular functional form for the event-by-event multiplicity fluctuations, choosing to use the one-parameter $\Gamma$ distribution (see Eq.~\eqref{eq:gammaflucs}) \cite{Moreland:2018gsh}. 

This choice restricts the range of models considered, excluding potentially viable models such as a CGC description of the initial state. In Ref.~\cite{Nagle:2018ybc} the authors investigated a CGC-like linear scaling of the initial energy density $\epsilon \propto T_A T_B$ and considered a log-normal distribution for the functional form of the multiplicity fluctuations. They found that, with appropriate choices of parameters and multiplicity fluctuations, these models can describe the data reasonably well.

It is important to study  the choice of multiplicity fluctuations and functional form in large and small systems. While it is well established that the QGP exists in large systems, signals of it have been found in the small system of pPb (ATLAS \cite{Chatrchyan:2013nka, Aaboud:2017acw, Aaboud:2017blb, Aad:2013fja}, CMS \cite{Sirunyan:2018toe, Chatrchyan:2013nka, Khachatryan:2014jra, Khachatryan:2015waa, Khachatryan:2015oea, Sirunyan:2017uyl}, and 
ALICE \cite{ABELEV:2013wsa, Abelev:2014mda}) and have been matched quantitatively by hydrodynamics \cite{Bozek:2011if, Bozek:2012gr, Bozek:2013ska, Bozek:2013uha, Kozlov:2014fqa, Zhou:2015iba, Zhao:2017rgg, Mantysaari:2017cni, Weller:2017tsr, Zhao:2017rgg}, though other explanations have been formulated \cite{Greif:2017bnr, Schenke:2016lrs, Mantysaari:2016ykx, Albacete:2017ajt}. Recently, a beam energy scan of $^3$HeAu and dAu by experimentalists at the RHIC PHENIX detector \cite{Aidala:2018mcw, Adare:2018toe} has also shown these systems to contain signs of the QGP. A recent analysis \cite{Schenke:2019pmk} looked at $v_2\{2\}$ in AuAu and dAu and found that precise measurements in these systems would be helpful in finding a signature  of the initial state momentum anisotropy.

Here, we systematically study the impact of these choices for the functional form of the entropy scaling and for multiplicity fluctuations on the initial state eccentricities. We find that there are identifiable differences between the two approaches that could be measured in experimental data through the fluctuations of $v_n$. 

\section{Methods}
Trento \cite{Moreland:2018gsh} constructs a nuclear profile function $T_A$ or $T_B$ through the formula $T_{A, B} = \omega_{A, B} \int dz \, \rho$, where $\rho$ is the number density of individual nucleons per unit volume and $\omega$ is a weight factor which fluctuates on an event-by-event basis.  The multiplicity weights introduce fluctuations from a distribution centered around $1$ and allows large single nucleon fluctuations. Trento assumes this distribution is a one-parameter $\Gamma$ distribution of the form 
\begin{equation} \label{eq:gammaflucs}
    P_{k}^{\Gamma}(\omega)=\frac{k^{k}}{\Gamma(k)}\omega^{k-1}e^{-k\omega},
\end{equation}
 where the shape of the distribution is controlled by $k$ \cite{Moreland:2018gsh}.
 \begin{figure}[ht]
    \begin{center}
    \includegraphics[width=0.48\textwidth]{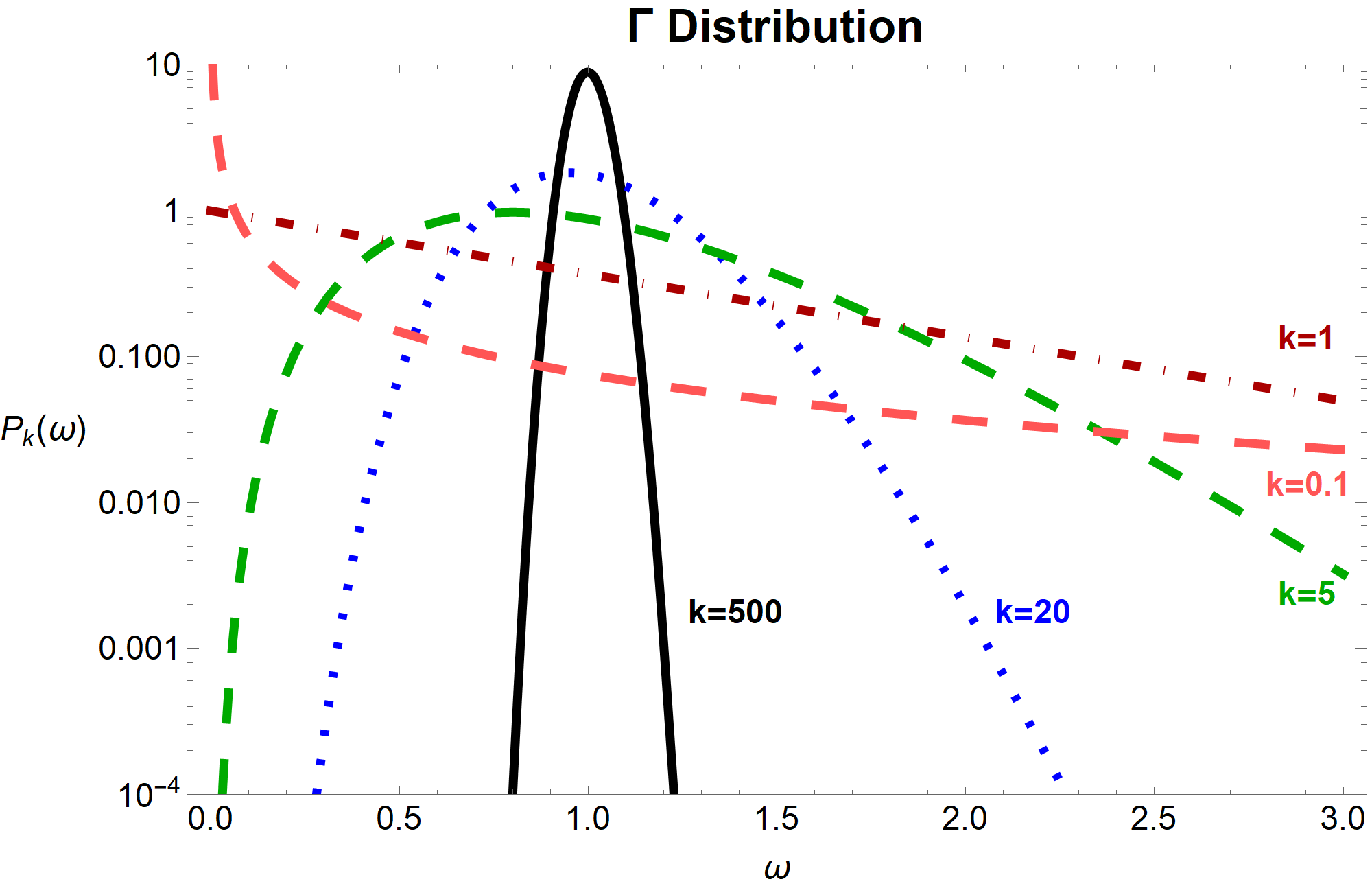}
    \includegraphics[width=0.48\textwidth]{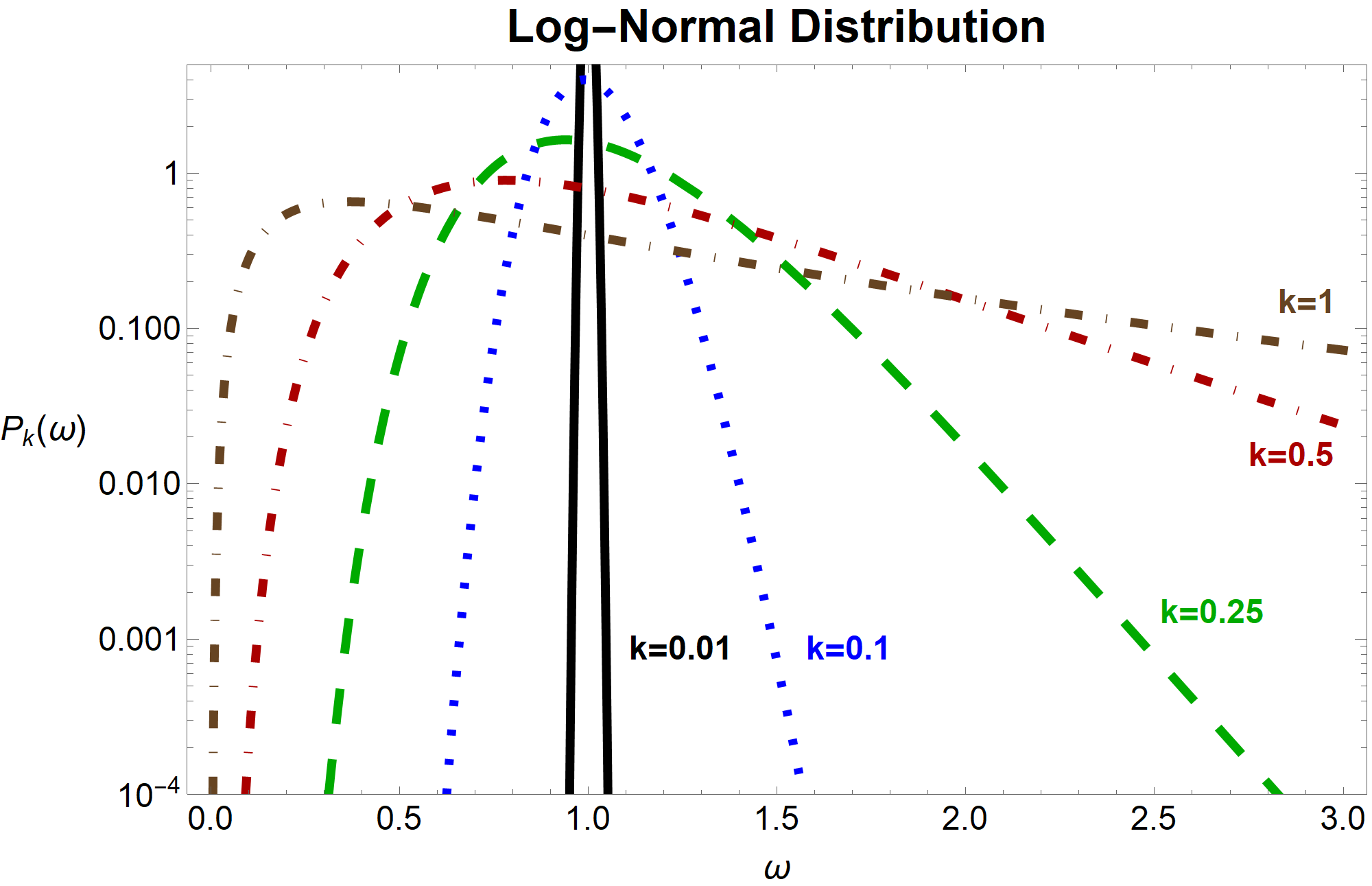}
    \caption{Probability distributions of gamma fluctuations Eq.\ \ref{eq:gammaflucs}, on the left, and log-normal Eq.\ \ref{eq:lognormflucs}, on the right. Several values of $k$ are shown to illustrate the the behaviour of the distributions.}
    \label{fig:flucdists}
    \end{center}
\end{figure}
The $\Gamma$ distribution is plotted in Fig. (\ref{fig:flucdists}) for several values of $k$. In the limit $k \rightarrow \infty$, the distribution Eq. (\ref{eq:gammaflucs}) approaches a delta function $\delta(\omega)$, and it becomes wider as $k\rightarrow1$.  On the other hand, when $k < 1$ a large divergence develops at $\omega = 0$ to counterbalance the extremely long large-$\omega$ tail. 

The assumption of the multiplicity fluctuations being determined by the $\Gamma$ distribution leads the Bayesian analysis to conclude the best initial state model is $p=0$, for which $s \propto \sqrt{T_{A}T_{B}}$. There are other models that have been shown to fit data that lie outside these assumptions and functional form, specifically linear scaling from CGC, $\epsilon \propto T_A T_B$, coupled with multiplicity fluctuations from a log-normal distribution
\begin{equation} \label{eq:lognormflucs}
    P_{k}^{log-normal}(\omega)=\frac{2}{\omega k\sqrt{2\pi}}e^{-\frac{\ln^{2}(\omega^{2})}{2k^{2}}}.
\end{equation}
The behaviour of this distribution is shown in Fig. (\ref{fig:flucdists}) for several values of $k$. In contrast to the $\Gamma$ distribution, the log-normal distribution Eq. (\ref{eq:lognormflucs}) approaches a delta function $\delta(\omega)$ when $k \rightarrow 0$ and widens for larger values $k \sim \mathcal{O}(1)$.  Unlike for the $\Gamma$ distribution, the log-normal distribution does not accumulate a peak at $\omega = 0$ for any value of $k$.

The initial state of heavy-ion collisions can be quantified by the eccentricity vectors $\mathcal{E}_n = \varepsilon_n e^{i\phi_n}$ where $\varepsilon_n$ is the magnitude and $\phi_n$ is the angle. Event eccentricities are calculated in the center-of-mass frame using the definition
\begin{equation}
    \mathcal{E}_n = - \frac{\int r^n e^{in\phi} f(r,\phi) rdrd\phi}
    {\int r^n f(r,\phi)rdrd\phi} = \varepsilon_n \, e^{i \phi_n}
\end{equation}
where the the function $f$ given in polar coordinates as $f(r,\phi)$ may be chosen to be either the energy density $f = \epsilon$ or the entropy density $f = s$.  In order to compare to the final state and in extension experimental data, two-particle cumulants \cite{Bilandzic:2010jr} of the eccentricity magnitudes are calculated using:
\begin{equation}
    \varepsilon_n\{2\}^2=\langle \varepsilon_n^2 \rangle
\end{equation}
where the subscript $n$ indicates the order of the eccentricity. We focus on $n = 2, 3$, which correspond to ellipticity and triangularity respectively, because they are dominated by linear response (higher harmonics have non-linear response already in their leading term \cite{Gardim:2011xv,Gardim:2014tya}). The approximate linear response  between the final-state flow harmonics $v_n$ and the initial-state eccentricities $\varepsilon_n$ can be described through
\begin{equation}
    v_n=\kappa_n \varepsilon_n,
\end{equation}
where $\kappa_n$ is a linear response coefficient.
 
The approximately linear response of $v_n$ to $\varepsilon_n$ is extremely useful, since it allows us to separate out medium effects by taking a ratio of cumulants. Ratios of multi-particle cumulants, specifically the ratio of the four-particle to two-particle cumulants, have been shown to provide important constraints on the initial state \cite{Giacalone:2017uqx}. The four-particle cumulant is defined as:
\begin{equation}    \label{e:var}
    \varepsilon_n\{4\}^4=2\langle \varepsilon_n^2 \rangle^2 - \langle \varepsilon_n^4 \rangle.
\end{equation}
Taking the ratio $\varepsilon_n\{4\}/\varepsilon_n\{2\}\/$ cancels out the response coefficient $\kappa_n$ and provides a direct comparison to $v_n\{4\}/v_n\{2\}$, which can be measured experimentally. As seen in \eqref{e:var}, this ratio also quantifies the amount of fluctuations in the quantity $v_n$ (or $\varepsilon_n$): the limit $\varepsilon_n \{4\} / \varepsilon_n \{2\} \rightarrow 1$ corresponds to no fluctuations ($\langle \varepsilon_n^2 \rangle^2 = \langle \varepsilon_n^4 \rangle$), with more fluctuations in $v_n$ or $\varepsilon_n$ the further below $1$ the cumulant ratio $\varepsilon_n \{4\} / \varepsilon_n \{2\}$ falls.

\section{Multiplicity distributions}

\begin{figure}[ht]
    \centering
    \includegraphics[width=0.48\textwidth]{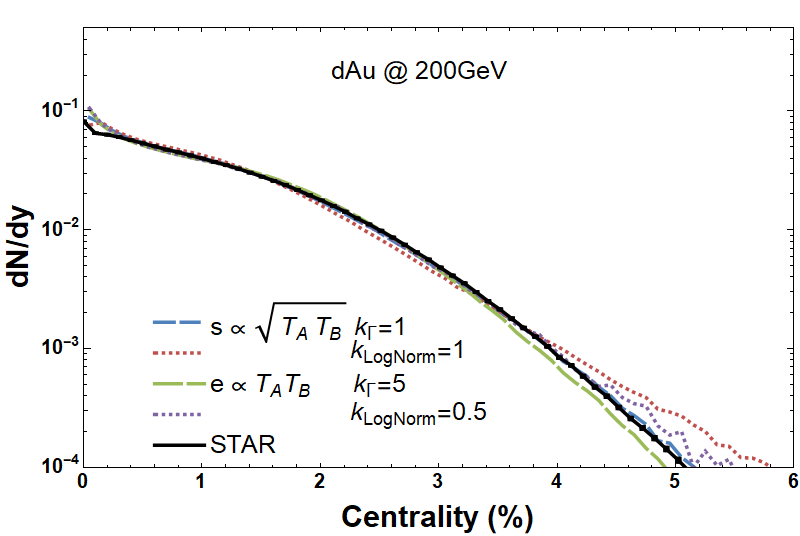}
    \caption{Multiplicity Distribution of dAu for functional forms $\sqrt{T_{A}T_{B}}$ and $T_{A}T_{B}$ using best fits for $\Gamma$ and lognormal multiplicity fluctuation distributions.}
    \label{fig:dAuMultiplicity}
\end{figure}

\begin{figure}[ht]
    \centering
    \includegraphics[width=0.48\textwidth]{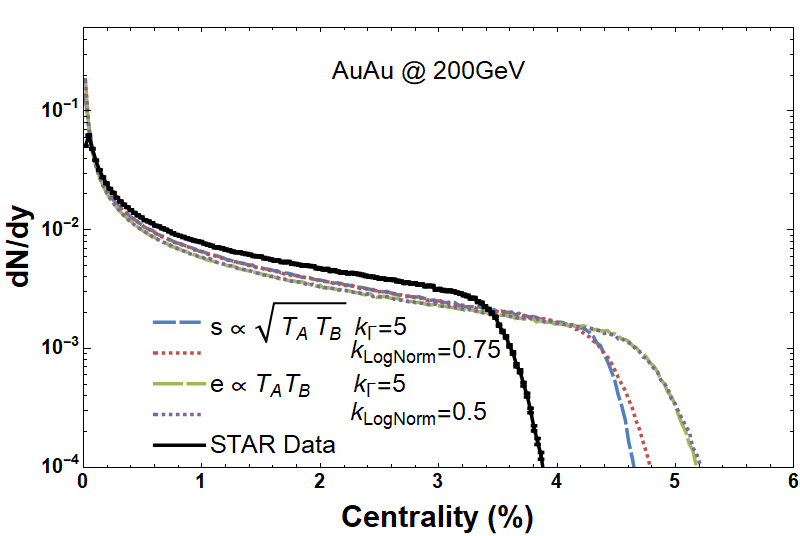}
    \caption{Multiplicity Distribution of AuAu for functional forms $\sqrt{T_{A}T_{B}}$ and $T_{A}T_{B}$ using best fits for $\Gamma$ and lognormal multiplicity fluctuation distributions.}
    \label{fig:AuAuMultiplicity}
\end{figure}

To study these effects, we have added the linear $T_A T_B$ scaling and log-normal fluctuations to Trento to make direct comparisons between the different scalings and fluctuations within the same model. Comparing experimental multiplicity distributions to estimates from the initial state requires the reasonable assumption that $dN/dy \propto S_0$, with $S_0$ being the initial total entropy of the event. We tune the parameter $k$ for each combination of functional form and multiplicity fluctuation distribution to best match STAR data for dAu and AuAu at 200 GeV \cite{Abelev:2008ab} in Figs. \ref{fig:dAuMultiplicity} and \ref{fig:AuAuMultiplicity}, respectively. 

For dAu the theoretical curves in Fig.\ \ref{fig:dAuMultiplicity} are all nearly identical below $N_{ch}/\langle N_{ch}\rangle < 4$ and, therefore, these differences in choices of the entropy scaling and multiplicity fluctuations only affect the multiplicity of dAu collisions in ultracentral events.  The linear $T_{A}T_{B}$ scaling is able to describe the ultracentral multiplicity tail for either the $\Gamma$ or log-normal distributions, with a similar amount of event-by-event fluctuations as compared to the phenomenological $\sqrt{T_{A}T_{B}}$. This implies that both scalings are capable of matching experimental data in dAu and there is no preference between them or the multiplicity fluctuations at this level.

For AuAu in Fig.\ \ref{fig:AuAuMultiplicity}, there is a slight suppression compared to experimental data below $N_{ch}/\langle N_{ch}\rangle = 3.5$. In this regime, the linear $T_A T_B$ scaling is slightly more suppressed than $\sqrt{T_A T_B}$ scaling.  For more central collisions neither entropy deposition model can correctly capture the high-multiplicity tail of the distribution. Both predict significantly wider fluctuations than what is seen by experiments. Generally, $\sqrt{T_{A}T_{B}}$ is somewhat closer to the STAR data than $T_{A}T_{B}$. We find little difference in our choice of gamma vs. log-normal fluctuations for this observable.

\begin{figure}[ht]
    \centering
    \includegraphics[width=0.48\textwidth]{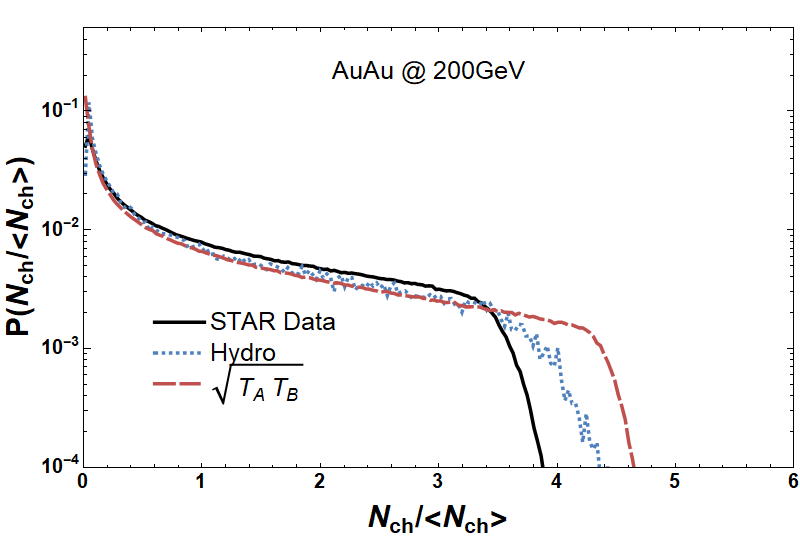}
    \caption{Illustration of the effect hydrodynamics has on the tail of the AuAu multiplicity distribution. }
    \label{fig:EffectOfHydro}
\end{figure}

The large difference between the tails of the experimental and initial-state multiplicity distributions in AuAu is troubling.  One possible remedy to this mismatch could be the effects of hydrodynamics on the multiplicity distribution. To explore this effect we use the hydrodynamic results from \cite{Alba:2017hhe,Rao:2019vgy} that coupled Trento+v-USPhydro \cite{Noronha-Hostler:2014dqa,Noronha-Hostler:2013gga} with a Lattice QCD based EOS \cite{Alba:2017hhe} and used the PDG16+ list \cite{Alba:2017mqu}. Additionally, only charged particles  are considered and kinematic cuts are implemented in the hydrodynamic results, as in the experiment. Comparing the multiplicity distribution for the $\sqrt{T_A T_B}$ scaling model before and after running hydrodynamics in Fig. \ref{fig:EffectOfHydro}, we see that the hydrodynamic evolution causes a significant decrease in the tail of the distribution. Hydrodynamic simulations  are expensive to run so there are far fewer events represented in the hydrodynamics curve than in the initial state curve, but putting a cut on the events for the initial state distribution does not change it and points to the dominant effect from the decrease in the tail being from hydrodynamics and not statistics.  While the corrections from hydrodynamics tend to move the $\sqrt{T_A T_B}$ multiplicity curves in the direction of the experimental data, they still do not fully resolve the mismatch.  Moreover, the linear scaling model $T_A T_B$ has a high-multiplicity tail which is even further away from the data than the $\sqrt{T_A T_B}$ case.  It is hard to see how such a model is consistent with the data, even after correcting the high-multiplicity tail with hydrodynamics.

\section{Eccentricities}

Returning to the argument of approximately linear response between the initial eccentricities and final flow harmonics, we explore the consequences of our choice in scaling and multiplicity fluctuations below.  We should note that in small systems linear response begins to break down \cite{Sievert:2019zjr,Schenke:2019pmk} so one should also explore these ideas in hydrodynamics, comparing to experimental data. However, due to the enormous cost of such simulations and current questions \cite{Plumberg:2021bme,Cheng:2021tnq} about causality in small systems, we leave that to a future work. Instead we focus solely on the eccentricities, which would have consequences for any Bayesian analyses regardless if linear or non-linear scaling \cite{Noronha-Hostler:2015dbi} occurs.

\subsection{AuAu}
\begin{figure}[ht]
    \begin{center}
    \includegraphics[width=0.48\textwidth]{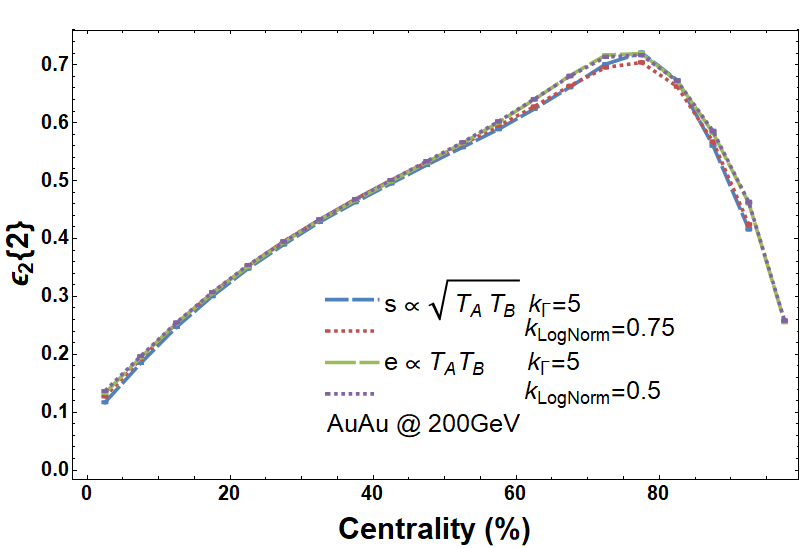}
    \includegraphics[width=0.48\textwidth]{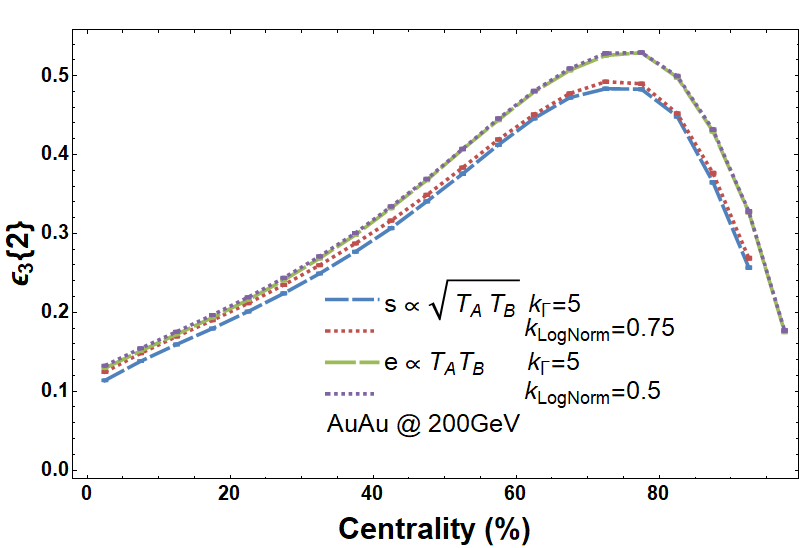}
    \caption{Two particle eccentricities of AuAu for functional forms $\sqrt{T_{A}T_{B}}$ and $T_{A}T_{B}$ using best fits for $\Gamma$ and lognormal multiplicity fluctuation distributions.}
    \label{fig:AuAueccs}
    \end{center}
\end{figure}

The influence of the choice in multiplicity fluctuations and entropy scaling model on the AuAu system are seen in the two-particle eccentricities in Fig. \ref{fig:AuAueccs}, $\varepsilon_{2}\{2\}$ (top) and $\varepsilon_{3}\{2\}$ (bottom). Below 60\% centrality in $\varepsilon_{2}\{2\}$ the two models agree, and there is little difference below 30\% centrality in $\varepsilon_{3}\{2\}$. The only significant difference between the two scaling models is seen in $\varepsilon_{3}\{2\}$ above 30\% centrality where linear has a larger magnitude than $\sqrt{T_A T_B}$. There is a slight spread with regard to the choice of multiplicity fluctuations in $\varepsilon_{3}\{2\}$ for the $\sqrt{T_A T_B}$ model in central collisions, but otherwise there is very little effect in this choice for both of the observables. 

\begin{figure}[ht]
    \begin{center}
    \includegraphics[width=0.48\textwidth]{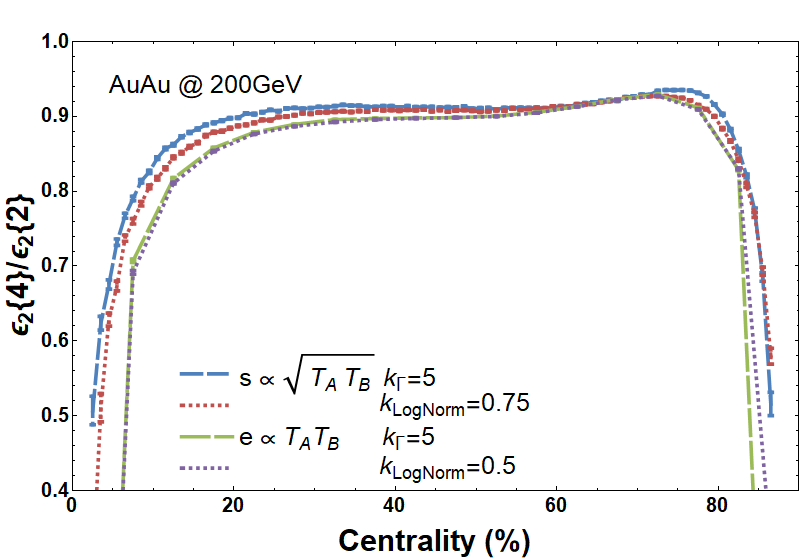}
    \includegraphics[width=0.48\textwidth]{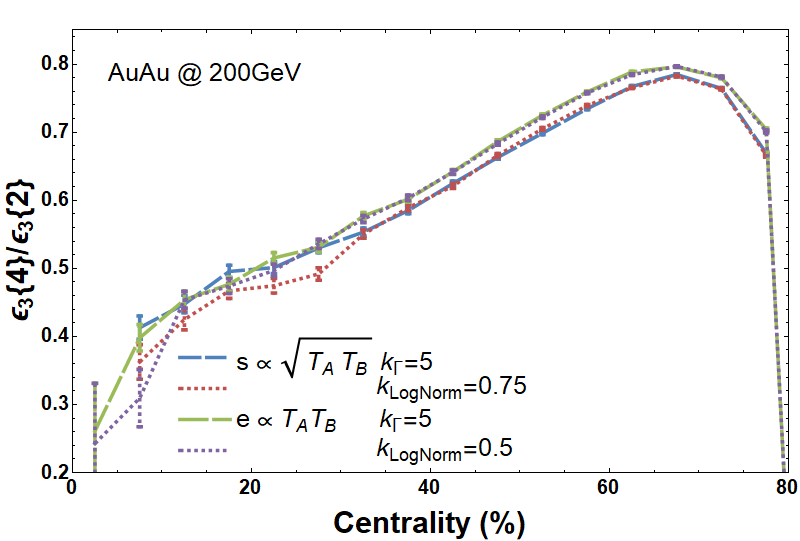}
    \caption{Four particle to two particle eccentricity ratios of AuAu for functional forms $\sqrt{T_{A}T_{B}}$ and $T_{A}T_{B}$ using best fits for $\Gamma$ and lognormal multiplicity fluctuation distributions.}
    \label{fig:AuAueccratios}
    \end{center}
\end{figure}

In Fig.~\ref{fig:AuAueccratios} we plot both  $\varepsilon_{2}\{4\}/\varepsilon_{2}\{2\}$ (top) and $\varepsilon_{3}\{4\}/\varepsilon_{3}\{2\}$ (bottom) vs. centrality comparing different scalings and multiplicity fluctuations. Looking at the four-particle to two-particle ratios for AuAu, the linear scaling $T_A T_B$ is lower in magnitude for $\varepsilon_{2}\{4\}/\varepsilon_{2}\{2\}$ in Fig. \ref{fig:AuAueccratios} (top) though it remains indifferent to choice of multiplicity fluctuation distribution. The $\sqrt{T_A T_B}$ scaling does show a separation in multiplicity fluctuation distribution, which may allow for a distinction to be made in comparison to the experimental data. However, one would require very precise experimental data to do so. It is also unsurprising that the largest differences occur in mid-central to central collisions because this is precisely where the effect of multiplicity fluctuations is the largest.

\begin{figure}[ht]
    \centering
    \includegraphics[width=0.48\textwidth]{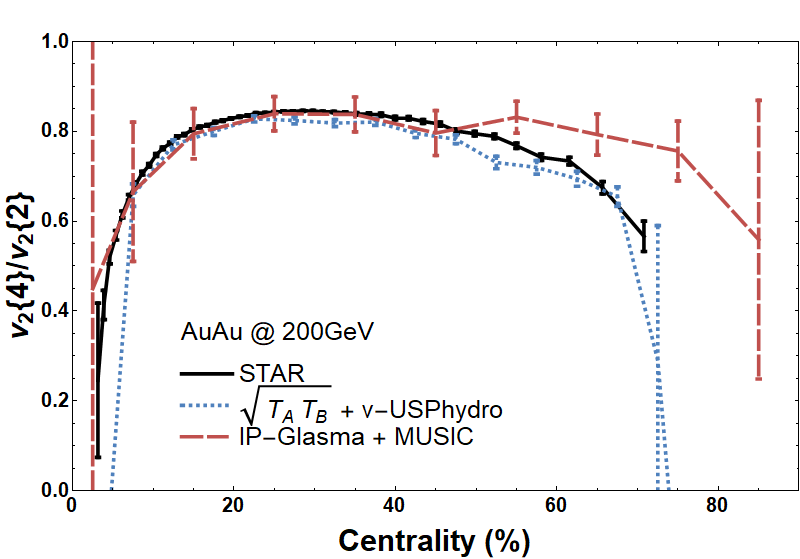}
    \caption{Comparison of the experimental measurement of $v_{2}\{4\}/v_{2}\{2\}$ from STAR \cite{STAR:2015mki} to two different hydrodynamic simulations \cite{Rao:2019vgy,Schenke:2019ruo} that reflect the different scaling models investigated in this paper.}
    \label{fig:EffectOfHydroOnV24V22}
\end{figure}

It is possible to estimate how this observable, $\varepsilon_{2}\{4\}/\varepsilon_{2}\{2\}$, will compare to experimental data by taking hydro simulations of similar scaling forms. In Fig. \ref{fig:EffectOfHydroOnV24V22}, the observable $v_{2}\{4\}/v_{2}\{2\}$ is plotted for the STAR experiment \cite{STAR:2015mki} and compared to hydro simulations using $\sqrt{T_A T_B}$ + v-USPhydro \cite{Rao:2019vgy}, to approximate the Trento preferred scaling, and IP-Glasma + MUSIC \cite{Schenke:2019ruo}, to approximate CGC-like scaling. Below 50\% centrality, both scalings match the data from STAR and above that there is a divergence in the CGC-like scaling. This difference at high centrality comes from non-linear response and is not expected to match. This comparison shows that the $\sqrt{T_A T_B}$ scaling from Trento and the CGC-like scaling are both able to match the most accurate data from STAR despite their differences in construction.

The case of $\varepsilon_{3}\{4\}/\varepsilon_{3}\{2\}$, in Fig. \ref{fig:AuAueccratios} (bottom) is shown.  We focus first on central collisions because this is precisely the regime where approximately linear response between $\varepsilon_{3}\{4\}/\varepsilon_{3}\{2\}$ and $v_{3}\{4\}/v_{3}\{2\}$ \cite{Carzon:2020xwp} works best.   In central collisions, the only discernible difference is the choice of multiplicity fluctuations between $\Gamma$ and log-normal distributions. The $\Gamma$ distribution produces fewer fluctuations than log-normal (larger values of $\varepsilon_{3}\{4\}/\varepsilon_{3}\{2\}$). In contrast, peripheral collisions (centralities $>30\%$) wash out any effects from the choice in multiplicity fluctuations and only depend on the choice of entropy scaling model, with linear $T_A T_B$ scaling having fewer fluctuations than $\sqrt{T_A T_B}$ scaling.

\subsection{dAu}
\begin{figure}[ht]
    \begin{center}
    \includegraphics[width=0.48\textwidth]{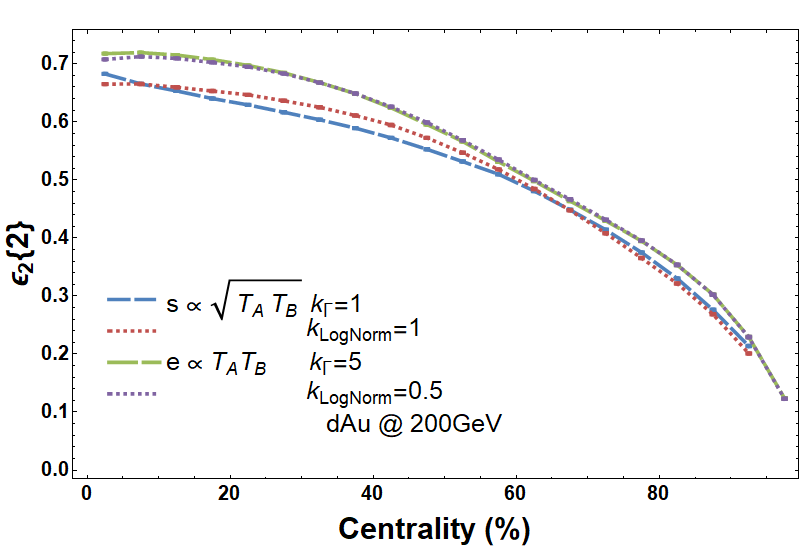}
    \includegraphics[width=0.48\textwidth]{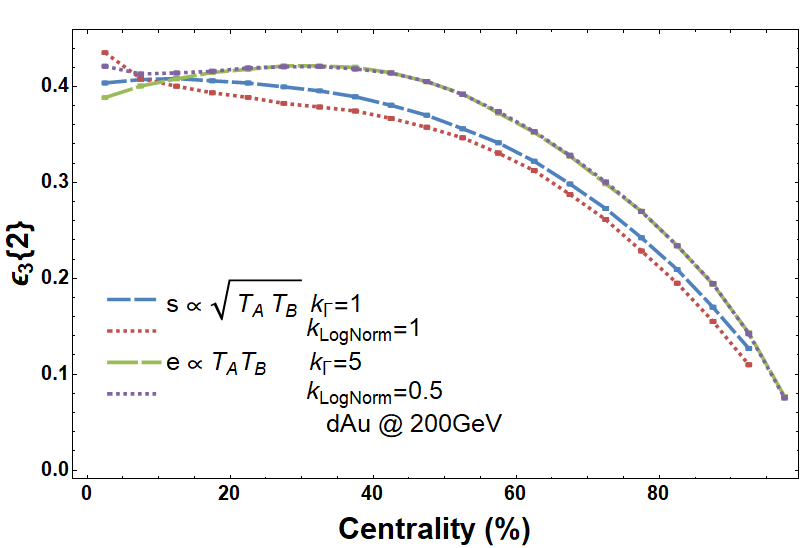}
    \caption{Two particle eccentricities of dAu for functional forms $\sqrt{T_{A}T_{B}}$ and $T_{A}T_{B}$ using best fits for $\Gamma$ and lognormal multiplicity fluctuation distributions.}
    \label{fig:dAueccs}
    \end{center}
\end{figure}

The effect of these choices on dAu collisions is more pronounced than in AuAu and can be seen in the plots of $\varepsilon_{2}\{2\}$ (top) and $\varepsilon_{3}\{2\}$ (bottom), shown in Fig. \ref{fig:dAueccs}.
The strongest sensitivity is to the choice of entropy deposition model. For all parameter choices and across both $\varepsilon_{2}\{2\}$ and $\varepsilon_{3}\{2\}$, the magnitude of the eccentricities for $T_{A}T_{B}$ is larger than for $\sqrt{T_A T_B}$ scaling, with an interesting reverse ordering in $\varepsilon_{3}\{2\}$ for central events. These differences in the eccentricities would lead to a different shear viscosity being extracted for both models to match the same experimental data.
The eccentricities of the $\sqrt{T_A T_B}$ model do not have much dependence on the choice of fluctuation distribution except for very central collisions, where the log-normal distribution stays flat in $\varepsilon_{2}\{2\}$ and curves up in $\varepsilon_{3}\{2\}$ whereas the $\Gamma$ distribution leads to an upward curve in $\varepsilon_{2}\{2\}$ and levels off in $\varepsilon_{3}\{2\}$.

\begin{figure}[ht]
    \begin{center}
    \includegraphics[width=0.48\textwidth]{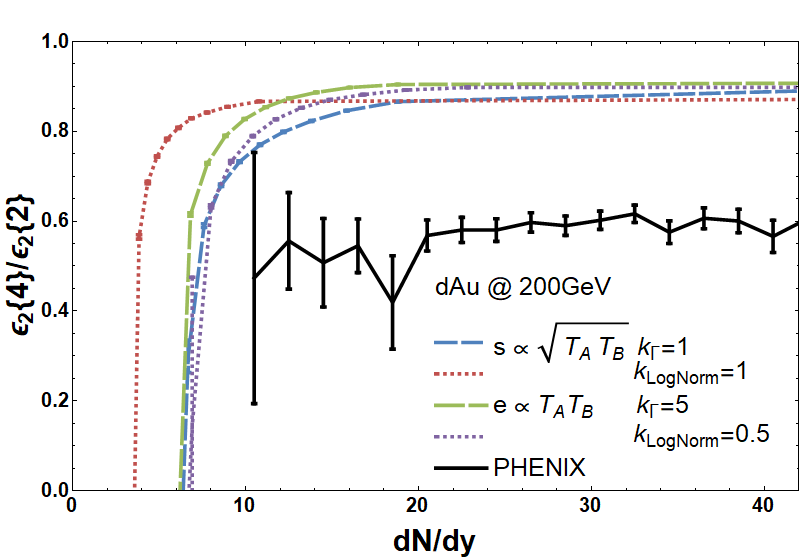}
    \includegraphics[width=0.48\textwidth]{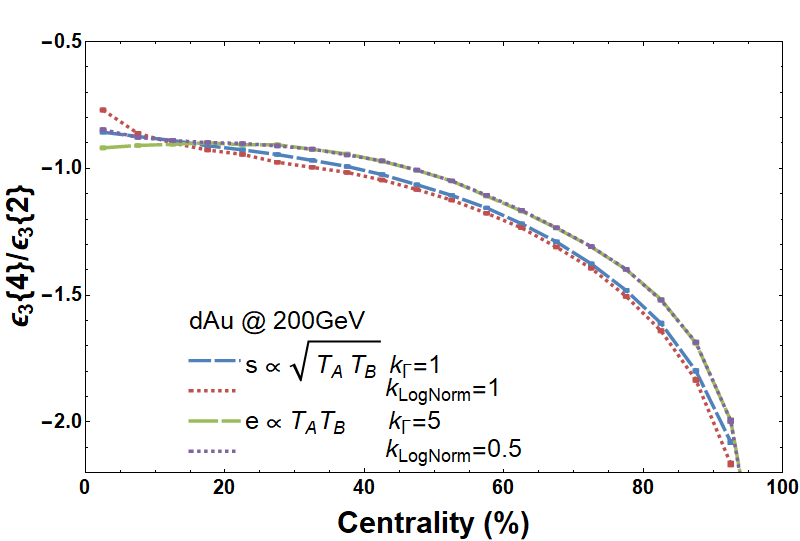}
    \caption{Four particle to two particle eccentricity ratios of dAu for functional forms $\sqrt{T_{A}T_{B}}$ and $T_{A}T_{B}$ using best fits for $\Gamma$ and lognormal multiplicity fluctuation distributions.  PHENIX data taken from Ref.~\cite{Aidala:2017ajz}.}
    \label{fig:dAueccratios}
    \end{center}
\end{figure}

A comparison of $\varepsilon_{2}\{4\}/\varepsilon_{2}\{2\}$ from PHENIX \cite{Aidala:2017ajz} to the different scaling models and multiplicity fluctuation distributions can be seen in Fig. \ref{fig:dAueccratios} (top). There is some ambiguity in the comparison of the experimental data from PHENIX that is plotted versus $N_{track}$ and the initial state calculation plotted versus $dN/dy$ (see e.g for more discussion on the topic \cite{ATLAS:2021kty}). Here we just assume that the two quantities are the same. 

There is a significant difference between the linear $T_A T_B$ and $\sqrt{T_A T_B}$ entropy scalings and a large effect from the choice of  multiplicity fluctuations in $\sqrt{T_{A}T_{B}}$ that was not seen in the two-particle eccentricities. The most useful feature is where the different parameter sets change sign, since this is at a different point for the $\sqrt{T_A T_B}$ model with either multiplicity distribution and the linear $T_A T_B$ scaling. It is difficult to draw specific conclusions from the comparison since the experimental data is dependent on final state properties that cannot be estimated by the initial state. A calculation of this ratio in terms of $dN/dy$ or percent centrality would be useful in determining where the sign change occurs and differentiating between the scaling models and even the multiplicity fluctuations for the $\sqrt{T_A T_B}$ form. 

In Fig. \ref{fig:dAueccratios} (bottom), $\varepsilon_{3}\{4\}/\varepsilon_{3}\{2\}$ is plotted for dAu. The trends from $\varepsilon_{3}\{2\}$ in Fig. \ref{fig:dAueccs} (bottom) are seen in the 4-particle to 2-particle ratio, with $T_{A}T_{B}$ having a higher magnitude than $\sqrt{T_{A}T_{B}}$ except at the most central events and log-normal fluctuations giving the $\sqrt{T_A T_B}$ form an upward curvature in central events while the $\Gamma$ distribution case remains flat. We note that here we chose to plot vs. centrality instead of $dN/dy$ because no experimental data yet exists for this measurement.

\section{Conclusion}
In Trento's Bayesian analysis, multiplicity fluctuations are assumed to be of the form of a $\Gamma$ distribution. This assumption may contribute to the preferred functional form $T_R = \sqrt{T_A T_B}$. Functional forms outside the scope of Trento's analysis, specifically linear scaling $T_R \propto T_A T_B$, are ignored, though they are able to match multiplicity distributions as well as the phenomenologically-preferred $\sqrt{T_A T_B}$ form in AuAu and dAu systems.  More precise implementations of the conversion of an initial energy density $\epsilon \propto T_A T_B$ and the corresponding initial entropy density $s$ will depend on the details of a particular equation of state.  These improvements can be made in future work, but the general effect is seen with the conformal conversion used here.  The assumption of multiplicity fluctuations following a $\Gamma$ distribution also excludes other distributions that have been used for this purpose, specifically log-normal. Adding linear $T_A T_B$ scaling and log-normal fluctuations to Trento, we see that the $\sqrt{T_A T_B}$ form does indeed prefer $\Gamma$ fluctuations while the linear functional form is able to match experimental data as well. While not explored in this paper because all simulations were boost invariant, multiplicity fluctuations in the longitudinal direction are also interesting \cite{Jia:2015jga,Jia:2020tvb} and could be studied in a future work. 

Across both AuAu and dAu, linear $T_A T_B$ has a higher magnitude in $\varepsilon_2 \{2\}$ and $\varepsilon_3 \{2\}$ than $\sqrt{T_A T_B}$ which may lead to the extraction of a different shear viscosity for the different functional forms, especially for small systems. Differences between models are enhanced in smaller systems and thus become more important when trying to understand them. These differences suggest that Trento's Bayesian extraction of QGP viscosities in small systems may contain a systematic uncertainty that could be controlled by increasing the allowed functional space. 

Ratios of $\varepsilon_{2}\{4\}/\varepsilon_{2}\{2\}$ and $\varepsilon_{3}\{4\}/\varepsilon_{3}\{2\}$ are important to look at as well since, due to linear response, they can be used to constrain parameters in the initial state with experiment. These ratios are important in dAu, where the sign change of $\varepsilon_{2}\{4\}/\varepsilon_{2}\{2\}$ could distinguish between models in the small system. While they may be costly to calculate in hydrodynamic models (when running a Bayesian analysis), one can obtain $v_{3}\{4\}/v_{3}\{2\}$ directly from eccentricities alone in ultracentral collisions (with less than $1\%$ error \cite{Carzon:2020xwp}).

\section*{Acknowledgements}
We would also like to thank Anthony Timmins, Guliano Giacalone, and Wilke van der Schee for discussions. P.C. and J.N.H. acknowledge support from the US-DOE Nuclear Science Grant No. DE-SC0020633 and the Illinois Campus Cluster, a computing resource that is operated by the Illinois Campus Cluster Program (ICCP) in conjunction with the National Center for Supercomputing Applications (NCSA), and which is supported by funds from the University of Illinois at Urbana-Champaign.  M.S. acknowledges support from a start-up grant from New Mexico State University.

\bibliography{bibliography}

\begin{thebibliography}{86}%
\makeatletter
\providecommand \@ifxundefined [1]{%
 \@ifx{#1\undefined}
}%
\providecommand \@ifnum [1]{%
 \ifnum #1\expandafter \@firstoftwo
 \else \expandafter \@secondoftwo
 \fi
}%
\providecommand \@ifx [1]{%
 \ifx #1\expandafter \@firstoftwo
 \else \expandafter \@secondoftwo
 \fi
}%
\providecommand \natexlab [1]{#1}%
\providecommand \enquote  [1]{``#1''}%
\providecommand \bibnamefont  [1]{#1}%
\providecommand \bibfnamefont [1]{#1}%
\providecommand \citenamefont [1]{#1}%
\providecommand \href@noop [0]{\@secondoftwo}%
\providecommand \href [0]{\begingroup \@sanitize@url \@href}%
\providecommand \@href[1]{\@@startlink{#1}\@@href}%
\providecommand \@@href[1]{\endgroup#1\@@endlink}%
\providecommand \@sanitize@url [0]{\catcode `\\12\catcode `\$12\catcode
  `\&12\catcode `\#12\catcode `\^12\catcode `\_12\catcode `\%12\relax}%
\providecommand \@@startlink[1]{}%
\providecommand \@@endlink[0]{}%
\providecommand \url  [0]{\begingroup\@sanitize@url \@url }%
\providecommand \@url [1]{\endgroup\@href {#1}{\urlprefix }}%
\providecommand \urlprefix  [0]{URL }%
\providecommand \Eprint [0]{\href }%
\providecommand \doibase [0]{http://dx.doi.org/}%
\providecommand \selectlanguage [0]{\@gobble}%
\providecommand \bibinfo  [0]{\@secondoftwo}%
\providecommand \bibfield  [0]{\@secondoftwo}%
\providecommand \translation [1]{[#1]}%
\providecommand \BibitemOpen [0]{}%
\providecommand \bibitemStop [0]{}%
\providecommand \bibitemNoStop [0]{.\EOS\space}%
\providecommand \EOS [0]{\spacefactor3000\relax}%
\providecommand \BibitemShut  [1]{\csname bibitem#1\endcsname}%
\let\auto@bib@innerbib\@empty
\bibitem [{\citenamefont {Noronha-Hostler}\ \emph
  {et~al.}(2016{\natexlab{a}})\citenamefont {Noronha-Hostler}, \citenamefont
  {Luzum},\ and\ \citenamefont {Ollitrault}}]{Noronha-Hostler:2015uye}%
  \BibitemOpen
  \bibfield  {author} {\bibinfo {author} {\bibfnamefont {J.}~\bibnamefont
  {Noronha-Hostler}}, \bibinfo {author} {\bibfnamefont {M.}~\bibnamefont
  {Luzum}}, \ and\ \bibinfo {author} {\bibfnamefont {J.-Y.}\ \bibnamefont
  {Ollitrault}},\ }\href {\doibase 10.1103/PhysRevC.93.034912} {\bibfield
  {journal} {\bibinfo  {journal} {Phys. Rev. C}\ }\textbf {\bibinfo {volume}
  {93}},\ \bibinfo {pages} {034912} (\bibinfo {year} {2016}{\natexlab{a}})},\
  \Eprint {http://arxiv.org/abs/1511.06289} {arXiv:1511.06289 [nucl-th]}
  \BibitemShut {NoStop}%
\bibitem [{\citenamefont {Niemi}\ \emph
  {et~al.}(2016{\natexlab{a}})\citenamefont {Niemi}, \citenamefont {Eskola},
  \citenamefont {Paatelainen},\ and\ \citenamefont {Tuominen}}]{Niemi:2015voa}%
  \BibitemOpen
  \bibfield  {author} {\bibinfo {author} {\bibfnamefont {H.}~\bibnamefont
  {Niemi}}, \bibinfo {author} {\bibfnamefont {K.~J.}\ \bibnamefont {Eskola}},
  \bibinfo {author} {\bibfnamefont {R.}~\bibnamefont {Paatelainen}}, \ and\
  \bibinfo {author} {\bibfnamefont {K.}~\bibnamefont {Tuominen}},\ }\href
  {\doibase 10.1103/PhysRevC.93.014912} {\bibfield  {journal} {\bibinfo
  {journal} {Phys. Rev. C}\ }\textbf {\bibinfo {volume} {93}},\ \bibinfo
  {pages} {014912} (\bibinfo {year} {2016}{\natexlab{a}})},\ \Eprint
  {http://arxiv.org/abs/1511.04296} {arXiv:1511.04296 [hep-ph]} \BibitemShut
  {NoStop}%
\bibitem [{\citenamefont {Adam}\ \emph {et~al.}(2016)\citenamefont {Adam} \emph
  {et~al.}}]{Adam:2015ptt}%
  \BibitemOpen
  \bibfield  {author} {\bibinfo {author} {\bibfnamefont {J.}~\bibnamefont
  {Adam}} \emph {et~al.} (\bibinfo {collaboration} {ALICE}),\ }\href {\doibase
  10.1103/PhysRevLett.116.222302} {\bibfield  {journal} {\bibinfo  {journal}
  {Phys. Rev. Lett.}\ }\textbf {\bibinfo {volume} {116}},\ \bibinfo {pages}
  {222302} (\bibinfo {year} {2016})},\ \Eprint
  {http://arxiv.org/abs/1512.06104} {arXiv:1512.06104 [nucl-ex]} \BibitemShut
  {NoStop}%
\bibitem [{\citenamefont {Song}\ \emph {et~al.}(2011)\citenamefont {Song},
  \citenamefont {Bass}, \citenamefont {Heinz}, \citenamefont {Hirano},\ and\
  \citenamefont {Shen}}]{Song:2010mg}%
  \BibitemOpen
  \bibfield  {author} {\bibinfo {author} {\bibfnamefont {H.}~\bibnamefont
  {Song}}, \bibinfo {author} {\bibfnamefont {S.~A.}\ \bibnamefont {Bass}},
  \bibinfo {author} {\bibfnamefont {U.}~\bibnamefont {Heinz}}, \bibinfo
  {author} {\bibfnamefont {T.}~\bibnamefont {Hirano}}, \ and\ \bibinfo {author}
  {\bibfnamefont {C.}~\bibnamefont {Shen}},\ }\href {\doibase
  10.1103/PhysRevLett.106.192301} {\bibfield  {journal} {\bibinfo  {journal}
  {Phys. Rev. Lett.}\ }\textbf {\bibinfo {volume} {106}},\ \bibinfo {pages}
  {192301} (\bibinfo {year} {2011})},\ \bibinfo {note} {[Erratum:
  Phys.Rev.Lett. 109, 139904 (2012)]},\ \Eprint
  {http://arxiv.org/abs/1011.2783} {arXiv:1011.2783 [nucl-th]} \BibitemShut
  {NoStop}%
\bibitem [{\citenamefont {Bozek}\ and\ \citenamefont
  {Wyskiel-Piekarska}(2012)}]{Bozek:2012qs}%
  \BibitemOpen
  \bibfield  {author} {\bibinfo {author} {\bibfnamefont {P.}~\bibnamefont
  {Bozek}}\ and\ \bibinfo {author} {\bibfnamefont {I.}~\bibnamefont
  {Wyskiel-Piekarska}},\ }\href {\doibase 10.1103/PhysRevC.85.064915}
  {\bibfield  {journal} {\bibinfo  {journal} {Phys. Rev. C}\ }\textbf {\bibinfo
  {volume} {85}},\ \bibinfo {pages} {064915} (\bibinfo {year} {2012})},\
  \Eprint {http://arxiv.org/abs/1203.6513} {arXiv:1203.6513 [nucl-th]}
  \BibitemShut {NoStop}%
\bibitem [{\citenamefont {Gardim}\ \emph
  {et~al.}(2012{\natexlab{a}})\citenamefont {Gardim}, \citenamefont {Grassi},
  \citenamefont {Luzum},\ and\ \citenamefont {Ollitrault}}]{Gardim:2012yp}%
  \BibitemOpen
  \bibfield  {author} {\bibinfo {author} {\bibfnamefont {F.~G.}\ \bibnamefont
  {Gardim}}, \bibinfo {author} {\bibfnamefont {F.}~\bibnamefont {Grassi}},
  \bibinfo {author} {\bibfnamefont {M.}~\bibnamefont {Luzum}}, \ and\ \bibinfo
  {author} {\bibfnamefont {J.-Y.}\ \bibnamefont {Ollitrault}},\ }\href
  {\doibase 10.1103/PhysRevLett.109.202302} {\bibfield  {journal} {\bibinfo
  {journal} {Phys. Rev. Lett.}\ }\textbf {\bibinfo {volume} {109}},\ \bibinfo
  {pages} {202302} (\bibinfo {year} {2012}{\natexlab{a}})},\ \Eprint
  {http://arxiv.org/abs/1203.2882} {arXiv:1203.2882 [nucl-th]} \BibitemShut
  {NoStop}%
\bibitem [{\citenamefont {Bozek}\ and\ \citenamefont
  {Broniowski}(2013{\natexlab{a}})}]{Bozek:2013uha}%
  \BibitemOpen
  \bibfield  {author} {\bibinfo {author} {\bibfnamefont {P.}~\bibnamefont
  {Bozek}}\ and\ \bibinfo {author} {\bibfnamefont {W.}~\bibnamefont
  {Broniowski}},\ }\href {\doibase 10.1103/PhysRevC.88.014903} {\bibfield
  {journal} {\bibinfo  {journal} {Phys. Rev. C}\ }\textbf {\bibinfo {volume}
  {88}},\ \bibinfo {pages} {014903} (\bibinfo {year} {2013}{\natexlab{a}})},\
  \Eprint {http://arxiv.org/abs/1304.3044} {arXiv:1304.3044 [nucl-th]}
  \BibitemShut {NoStop}%
\bibitem [{\citenamefont {Niemi}\ \emph
  {et~al.}(2016{\natexlab{b}})\citenamefont {Niemi}, \citenamefont {Eskola},\
  and\ \citenamefont {Paatelainen}}]{Niemi:2015qia}%
  \BibitemOpen
  \bibfield  {author} {\bibinfo {author} {\bibfnamefont {H.}~\bibnamefont
  {Niemi}}, \bibinfo {author} {\bibfnamefont {K.~J.}\ \bibnamefont {Eskola}}, \
  and\ \bibinfo {author} {\bibfnamefont {R.}~\bibnamefont {Paatelainen}},\
  }\href {\doibase 10.1103/PhysRevC.93.024907} {\bibfield  {journal} {\bibinfo
  {journal} {Phys. Rev. C}\ }\textbf {\bibinfo {volume} {93}},\ \bibinfo
  {pages} {024907} (\bibinfo {year} {2016}{\natexlab{b}})},\ \Eprint
  {http://arxiv.org/abs/1505.02677} {arXiv:1505.02677 [hep-ph]} \BibitemShut
  {NoStop}%
\bibitem [{\citenamefont {Ryu}\ \emph {et~al.}(2015)\citenamefont {Ryu},
  \citenamefont {Paquet}, \citenamefont {Shen}, \citenamefont {Denicol},
  \citenamefont {Schenke}, \citenamefont {Jeon},\ and\ \citenamefont
  {Gale}}]{Ryu:2015vwa}%
  \BibitemOpen
  \bibfield  {author} {\bibinfo {author} {\bibfnamefont {S.}~\bibnamefont
  {Ryu}}, \bibinfo {author} {\bibfnamefont {J.~F.}\ \bibnamefont {Paquet}},
  \bibinfo {author} {\bibfnamefont {C.}~\bibnamefont {Shen}}, \bibinfo {author}
  {\bibfnamefont {G.~S.}\ \bibnamefont {Denicol}}, \bibinfo {author}
  {\bibfnamefont {B.}~\bibnamefont {Schenke}}, \bibinfo {author} {\bibfnamefont
  {S.}~\bibnamefont {Jeon}}, \ and\ \bibinfo {author} {\bibfnamefont
  {C.}~\bibnamefont {Gale}},\ }\href {\doibase 10.1103/PhysRevLett.115.132301}
  {\bibfield  {journal} {\bibinfo  {journal} {Phys. Rev. Lett.}\ }\textbf
  {\bibinfo {volume} {115}},\ \bibinfo {pages} {132301} (\bibinfo {year}
  {2015})},\ \Eprint {http://arxiv.org/abs/1502.01675} {arXiv:1502.01675
  [nucl-th]} \BibitemShut {NoStop}%
\bibitem [{\citenamefont {McDonald}\ \emph {et~al.}(2017)\citenamefont
  {McDonald}, \citenamefont {Shen}, \citenamefont {Fillion-Gourdeau},
  \citenamefont {Jeon},\ and\ \citenamefont {Gale}}]{McDonald:2016vlt}%
  \BibitemOpen
  \bibfield  {author} {\bibinfo {author} {\bibfnamefont {S.}~\bibnamefont
  {McDonald}}, \bibinfo {author} {\bibfnamefont {C.}~\bibnamefont {Shen}},
  \bibinfo {author} {\bibfnamefont {F.}~\bibnamefont {Fillion-Gourdeau}},
  \bibinfo {author} {\bibfnamefont {S.}~\bibnamefont {Jeon}}, \ and\ \bibinfo
  {author} {\bibfnamefont {C.}~\bibnamefont {Gale}},\ }\href {\doibase
  10.1103/PhysRevC.95.064913} {\bibfield  {journal} {\bibinfo  {journal} {Phys.
  Rev. C}\ }\textbf {\bibinfo {volume} {95}},\ \bibinfo {pages} {064913}
  (\bibinfo {year} {2017})},\ \Eprint {http://arxiv.org/abs/1609.02958}
  {arXiv:1609.02958 [hep-ph]} \BibitemShut {NoStop}%
\bibitem [{\citenamefont {Bernhard}\ \emph {et~al.}(2016)\citenamefont
  {Bernhard}, \citenamefont {Moreland}, \citenamefont {Bass}, \citenamefont
  {Liu},\ and\ \citenamefont {Heinz}}]{Bernhard:2016tnd}%
  \BibitemOpen
  \bibfield  {author} {\bibinfo {author} {\bibfnamefont {J.~E.}\ \bibnamefont
  {Bernhard}}, \bibinfo {author} {\bibfnamefont {J.~S.}\ \bibnamefont
  {Moreland}}, \bibinfo {author} {\bibfnamefont {S.~A.}\ \bibnamefont {Bass}},
  \bibinfo {author} {\bibfnamefont {J.}~\bibnamefont {Liu}}, \ and\ \bibinfo
  {author} {\bibfnamefont {U.}~\bibnamefont {Heinz}},\ }\href {\doibase
  10.1103/PhysRevC.94.024907} {\bibfield  {journal} {\bibinfo  {journal} {Phys.
  Rev. C}\ }\textbf {\bibinfo {volume} {94}},\ \bibinfo {pages} {024907}
  (\bibinfo {year} {2016})},\ \Eprint {http://arxiv.org/abs/1605.03954}
  {arXiv:1605.03954 [nucl-th]} \BibitemShut {NoStop}%
\bibitem [{\citenamefont {Gardim}\ \emph {et~al.}(2017)\citenamefont {Gardim},
  \citenamefont {Grassi}, \citenamefont {Luzum},\ and\ \citenamefont
  {Noronha-Hostler}}]{Gardim:2016nrr}%
  \BibitemOpen
  \bibfield  {author} {\bibinfo {author} {\bibfnamefont {F.~G.}\ \bibnamefont
  {Gardim}}, \bibinfo {author} {\bibfnamefont {F.}~\bibnamefont {Grassi}},
  \bibinfo {author} {\bibfnamefont {M.}~\bibnamefont {Luzum}}, \ and\ \bibinfo
  {author} {\bibfnamefont {J.}~\bibnamefont {Noronha-Hostler}},\ }\href
  {\doibase 10.1103/PhysRevC.95.034901} {\bibfield  {journal} {\bibinfo
  {journal} {Phys. Rev. C}\ }\textbf {\bibinfo {volume} {95}},\ \bibinfo
  {pages} {034901} (\bibinfo {year} {2017})},\ \Eprint
  {http://arxiv.org/abs/1608.02982} {arXiv:1608.02982 [nucl-th]} \BibitemShut
  {NoStop}%
\bibitem [{\citenamefont {Alba}\ \emph {et~al.}(2018)\citenamefont {Alba},
  \citenamefont {Mantovani~Sarti}, \citenamefont {Noronha}, \citenamefont
  {Noronha-Hostler}, \citenamefont {Parotto}, \citenamefont
  {Portillo~Vazquez},\ and\ \citenamefont {Ratti}}]{Alba:2017hhe}%
  \BibitemOpen
  \bibfield  {author} {\bibinfo {author} {\bibfnamefont {P.}~\bibnamefont
  {Alba}}, \bibinfo {author} {\bibfnamefont {V.}~\bibnamefont
  {Mantovani~Sarti}}, \bibinfo {author} {\bibfnamefont {J.}~\bibnamefont
  {Noronha}}, \bibinfo {author} {\bibfnamefont {J.}~\bibnamefont
  {Noronha-Hostler}}, \bibinfo {author} {\bibfnamefont {P.}~\bibnamefont
  {Parotto}}, \bibinfo {author} {\bibfnamefont {I.}~\bibnamefont
  {Portillo~Vazquez}}, \ and\ \bibinfo {author} {\bibfnamefont
  {C.}~\bibnamefont {Ratti}},\ }\href {\doibase 10.1103/PhysRevC.98.034909}
  {\bibfield  {journal} {\bibinfo  {journal} {Phys. Rev. C}\ }\textbf {\bibinfo
  {volume} {98}},\ \bibinfo {pages} {034909} (\bibinfo {year} {2018})},\
  \Eprint {http://arxiv.org/abs/1711.05207} {arXiv:1711.05207 [nucl-th]}
  \BibitemShut {NoStop}%
\bibitem [{\citenamefont {Giacalone}\ \emph {et~al.}(2018)\citenamefont
  {Giacalone}, \citenamefont {Noronha-Hostler}, \citenamefont {Luzum},\ and\
  \citenamefont {Ollitrault}}]{Giacalone:2017dud}%
  \BibitemOpen
  \bibfield  {author} {\bibinfo {author} {\bibfnamefont {G.}~\bibnamefont
  {Giacalone}}, \bibinfo {author} {\bibfnamefont {J.}~\bibnamefont
  {Noronha-Hostler}}, \bibinfo {author} {\bibfnamefont {M.}~\bibnamefont
  {Luzum}}, \ and\ \bibinfo {author} {\bibfnamefont {J.-Y.}\ \bibnamefont
  {Ollitrault}},\ }\href {\doibase 10.1103/PhysRevC.97.034904} {\bibfield
  {journal} {\bibinfo  {journal} {Phys. Rev. C}\ }\textbf {\bibinfo {volume}
  {97}},\ \bibinfo {pages} {034904} (\bibinfo {year} {2018})},\ \Eprint
  {http://arxiv.org/abs/1711.08499} {arXiv:1711.08499 [nucl-th]} \BibitemShut
  {NoStop}%
\bibitem [{\citenamefont {Eskola}\ \emph {et~al.}(2018)\citenamefont {Eskola},
  \citenamefont {Niemi}, \citenamefont {Paatelainen},\ and\ \citenamefont
  {Tuominen}}]{Eskola:2017bup}%
  \BibitemOpen
  \bibfield  {author} {\bibinfo {author} {\bibfnamefont {K.~J.}\ \bibnamefont
  {Eskola}}, \bibinfo {author} {\bibfnamefont {H.}~\bibnamefont {Niemi}},
  \bibinfo {author} {\bibfnamefont {R.}~\bibnamefont {Paatelainen}}, \ and\
  \bibinfo {author} {\bibfnamefont {K.}~\bibnamefont {Tuominen}},\ }\href
  {\doibase 10.1103/PhysRevC.97.034911} {\bibfield  {journal} {\bibinfo
  {journal} {Phys. Rev. C}\ }\textbf {\bibinfo {volume} {97}},\ \bibinfo
  {pages} {034911} (\bibinfo {year} {2018})},\ \Eprint
  {http://arxiv.org/abs/1711.09803} {arXiv:1711.09803 [hep-ph]} \BibitemShut
  {NoStop}%
\bibitem [{\citenamefont {Weller}\ and\ \citenamefont
  {Romatschke}(2017)}]{Weller:2017tsr}%
  \BibitemOpen
  \bibfield  {author} {\bibinfo {author} {\bibfnamefont {R.~D.}\ \bibnamefont
  {Weller}}\ and\ \bibinfo {author} {\bibfnamefont {P.}~\bibnamefont
  {Romatschke}},\ }\href {\doibase 10.1016/j.physletb.2017.09.077} {\bibfield
  {journal} {\bibinfo  {journal} {Phys. Lett. B}\ }\textbf {\bibinfo {volume}
  {774}},\ \bibinfo {pages} {351} (\bibinfo {year} {2017})},\ \Eprint
  {http://arxiv.org/abs/1701.07145} {arXiv:1701.07145 [nucl-th]} \BibitemShut
  {NoStop}%
\bibitem [{\citenamefont {Schenke}\ \emph {et~al.}(2019)\citenamefont
  {Schenke}, \citenamefont {Shen},\ and\ \citenamefont
  {Tribedy}}]{Schenke:2019ruo}%
  \BibitemOpen
  \bibfield  {author} {\bibinfo {author} {\bibfnamefont {B.}~\bibnamefont
  {Schenke}}, \bibinfo {author} {\bibfnamefont {C.}~\bibnamefont {Shen}}, \
  and\ \bibinfo {author} {\bibfnamefont {P.}~\bibnamefont {Tribedy}},\ }\href
  {\doibase 10.1103/PhysRevC.99.044908} {\bibfield  {journal} {\bibinfo
  {journal} {Phys. Rev. C}\ }\textbf {\bibinfo {volume} {99}},\ \bibinfo
  {pages} {044908} (\bibinfo {year} {2019})},\ \Eprint
  {http://arxiv.org/abs/1901.04378} {arXiv:1901.04378 [nucl-th]} \BibitemShut
  {NoStop}%
\bibitem [{\citenamefont {Gardim}\ \emph
  {et~al.}(2012{\natexlab{b}})\citenamefont {Gardim}, \citenamefont {Grassi},
  \citenamefont {Luzum},\ and\ \citenamefont {Ollitrault}}]{Gardim:2011xv}%
  \BibitemOpen
  \bibfield  {author} {\bibinfo {author} {\bibfnamefont {F.~G.}\ \bibnamefont
  {Gardim}}, \bibinfo {author} {\bibfnamefont {F.}~\bibnamefont {Grassi}},
  \bibinfo {author} {\bibfnamefont {M.}~\bibnamefont {Luzum}}, \ and\ \bibinfo
  {author} {\bibfnamefont {J.-Y.}\ \bibnamefont {Ollitrault}},\ }\href
  {\doibase 10.1103/PhysRevC.85.024908} {\bibfield  {journal} {\bibinfo
  {journal} {Phys. Rev. C}\ }\textbf {\bibinfo {volume} {85}},\ \bibinfo
  {pages} {024908} (\bibinfo {year} {2012}{\natexlab{b}})},\ \Eprint
  {http://arxiv.org/abs/1111.6538} {arXiv:1111.6538 [nucl-th]} \BibitemShut
  {NoStop}%
\bibitem [{\citenamefont {Gardim}\ \emph {et~al.}(2015)\citenamefont {Gardim},
  \citenamefont {Noronha-Hostler}, \citenamefont {Luzum},\ and\ \citenamefont
  {Grassi}}]{Gardim:2014tya}%
  \BibitemOpen
  \bibfield  {author} {\bibinfo {author} {\bibfnamefont {F.~G.}\ \bibnamefont
  {Gardim}}, \bibinfo {author} {\bibfnamefont {J.}~\bibnamefont
  {Noronha-Hostler}}, \bibinfo {author} {\bibfnamefont {M.}~\bibnamefont
  {Luzum}}, \ and\ \bibinfo {author} {\bibfnamefont {F.}~\bibnamefont
  {Grassi}},\ }\href {\doibase 10.1103/PhysRevC.91.034902} {\bibfield
  {journal} {\bibinfo  {journal} {Phys. Rev. C}\ }\textbf {\bibinfo {volume}
  {91}},\ \bibinfo {pages} {034902} (\bibinfo {year} {2015})},\ \Eprint
  {http://arxiv.org/abs/1411.2574} {arXiv:1411.2574 [nucl-th]} \BibitemShut
  {NoStop}%
\bibitem [{\citenamefont {Luzum}\ and\ \citenamefont
  {Romatschke}(2009)}]{Luzum:2009sb}%
  \BibitemOpen
  \bibfield  {author} {\bibinfo {author} {\bibfnamefont {M.}~\bibnamefont
  {Luzum}}\ and\ \bibinfo {author} {\bibfnamefont {P.}~\bibnamefont
  {Romatschke}},\ }\href {\doibase 10.1103/PhysRevLett.103.262302} {\bibfield
  {journal} {\bibinfo  {journal} {Phys. Rev. Lett.}\ }\textbf {\bibinfo
  {volume} {103}},\ \bibinfo {pages} {262302} (\bibinfo {year} {2009})},\
  \Eprint {http://arxiv.org/abs/0901.4588} {arXiv:0901.4588 [nucl-th]}
  \BibitemShut {NoStop}%
\bibitem [{\citenamefont {Bernhard}\ \emph {et~al.}(2015)\citenamefont
  {Bernhard}, \citenamefont {Marcy}, \citenamefont {Coleman-Smith},
  \citenamefont {Huzurbazar}, \citenamefont {Wolpert},\ and\ \citenamefont
  {Bass}}]{Bernhard:2015hxa}%
  \BibitemOpen
  \bibfield  {author} {\bibinfo {author} {\bibfnamefont {J.~E.}\ \bibnamefont
  {Bernhard}}, \bibinfo {author} {\bibfnamefont {P.~W.}\ \bibnamefont {Marcy}},
  \bibinfo {author} {\bibfnamefont {C.~E.}\ \bibnamefont {Coleman-Smith}},
  \bibinfo {author} {\bibfnamefont {S.}~\bibnamefont {Huzurbazar}}, \bibinfo
  {author} {\bibfnamefont {R.~L.}\ \bibnamefont {Wolpert}}, \ and\ \bibinfo
  {author} {\bibfnamefont {S.~A.}\ \bibnamefont {Bass}},\ }\href {\doibase
  10.1103/PhysRevC.91.054910} {\bibfield  {journal} {\bibinfo  {journal} {Phys.
  Rev. C}\ }\textbf {\bibinfo {volume} {91}},\ \bibinfo {pages} {054910}
  (\bibinfo {year} {2015})},\ \Eprint {http://arxiv.org/abs/1502.00339}
  {arXiv:1502.00339 [nucl-th]} \BibitemShut {NoStop}%
\bibitem [{\citenamefont {Heinz}\ \emph {et~al.}(2012)\citenamefont {Heinz},
  \citenamefont {Shen},\ and\ \citenamefont {Song}}]{Heinz:2011kt}%
  \BibitemOpen
  \bibfield  {author} {\bibinfo {author} {\bibfnamefont {U.}~\bibnamefont
  {Heinz}}, \bibinfo {author} {\bibfnamefont {C.}~\bibnamefont {Shen}}, \ and\
  \bibinfo {author} {\bibfnamefont {H.}~\bibnamefont {Song}},\ }\href {\doibase
  10.1063/1.3700674} {\bibfield  {journal} {\bibinfo  {journal} {AIP Conf.
  Proc.}\ }\textbf {\bibinfo {volume} {1441}},\ \bibinfo {pages} {766}
  (\bibinfo {year} {2012})},\ \Eprint {http://arxiv.org/abs/1108.5323}
  {arXiv:1108.5323 [nucl-th]} \BibitemShut {NoStop}%
\bibitem [{\citenamefont {Kharzeev}\ and\ \citenamefont
  {Nardi}(2001)}]{Kharzeev:2000ph}%
  \BibitemOpen
  \bibfield  {author} {\bibinfo {author} {\bibfnamefont {D.}~\bibnamefont
  {Kharzeev}}\ and\ \bibinfo {author} {\bibfnamefont {M.}~\bibnamefont
  {Nardi}},\ }\href {\doibase 10.1016/S0370-2693(01)00457-9} {\bibfield
  {journal} {\bibinfo  {journal} {Phys. Lett. B}\ }\textbf {\bibinfo {volume}
  {507}},\ \bibinfo {pages} {121} (\bibinfo {year} {2001})},\ \Eprint
  {http://arxiv.org/abs/nucl-th/0012025} {arXiv:nucl-th/0012025} \BibitemShut
  {NoStop}%
\bibitem [{\citenamefont {Goldschmidt}\ \emph {et~al.}(2015)\citenamefont
  {Goldschmidt}, \citenamefont {Qiu}, \citenamefont {Shen},\ and\ \citenamefont
  {Heinz}}]{Goldschmidt:2015kpa}%
  \BibitemOpen
  \bibfield  {author} {\bibinfo {author} {\bibfnamefont {A.}~\bibnamefont
  {Goldschmidt}}, \bibinfo {author} {\bibfnamefont {Z.}~\bibnamefont {Qiu}},
  \bibinfo {author} {\bibfnamefont {C.}~\bibnamefont {Shen}}, \ and\ \bibinfo
  {author} {\bibfnamefont {U.}~\bibnamefont {Heinz}},\ }\href {\doibase
  10.1103/PhysRevC.92.044903} {\bibfield  {journal} {\bibinfo  {journal} {Phys.
  Rev. C}\ }\textbf {\bibinfo {volume} {92}},\ \bibinfo {pages} {044903}
  (\bibinfo {year} {2015})},\ \Eprint {http://arxiv.org/abs/1507.03910}
  {arXiv:1507.03910 [nucl-th]} \BibitemShut {NoStop}%
\bibitem [{\citenamefont {Voloshin}(2010)}]{Voloshin:2010ut}%
  \BibitemOpen
  \bibfield  {author} {\bibinfo {author} {\bibfnamefont {S.~A.}\ \bibnamefont
  {Voloshin}},\ }\href {\doibase 10.1103/PhysRevLett.105.172301} {\bibfield
  {journal} {\bibinfo  {journal} {Phys. Rev. Lett.}\ }\textbf {\bibinfo
  {volume} {105}},\ \bibinfo {pages} {172301} (\bibinfo {year} {2010})},\
  \Eprint {http://arxiv.org/abs/1006.1020} {arXiv:1006.1020 [nucl-th]}
  \BibitemShut {NoStop}%
\bibitem [{\citenamefont {Pandit}(2013)}]{Pandit:2013uiv}%
  \BibitemOpen
  \bibfield  {author} {\bibinfo {author} {\bibfnamefont {Y.}~\bibnamefont
  {Pandit}} (\bibinfo {collaboration} {STAR}),\ }\href {\doibase
  10.1088/1742-6596/458/1/012003} {\bibfield  {journal} {\bibinfo  {journal}
  {J. Phys. Conf. Ser.}\ }\textbf {\bibinfo {volume} {458}},\ \bibinfo {pages}
  {012003} (\bibinfo {year} {2013})},\ \Eprint {http://arxiv.org/abs/1305.0173}
  {arXiv:1305.0173 [nucl-ex]} \BibitemShut {NoStop}%
\bibitem [{\citenamefont {Wang}\ and\ \citenamefont
  {Sorensen}(2014)}]{wang:2014qxa}%
  \BibitemOpen
  \bibfield  {author} {\bibinfo {author} {\bibfnamefont {H.}~\bibnamefont
  {Wang}}\ and\ \bibinfo {author} {\bibfnamefont {P.}~\bibnamefont {Sorensen}}
  (\bibinfo {collaboration} {STAR}),\ }\href {\doibase
  10.1016/j.nuclphysa.2014.09.111} {\bibfield  {journal} {\bibinfo  {journal}
  {Nucl. Phys. A}\ }\textbf {\bibinfo {volume} {932}},\ \bibinfo {pages} {169}
  (\bibinfo {year} {2014})},\ \Eprint {http://arxiv.org/abs/1406.7522}
  {arXiv:1406.7522 [nucl-ex]} \BibitemShut {NoStop}%
\bibitem [{\citenamefont {Adamczyk}\ \emph
  {et~al.}(2015{\natexlab{a}})\citenamefont {Adamczyk} \emph
  {et~al.}}]{adamczyk:2015obl}%
  \BibitemOpen
  \bibfield  {author} {\bibinfo {author} {\bibfnamefont {L.}~\bibnamefont
  {Adamczyk}} \emph {et~al.} (\bibinfo {collaboration} {STAR}),\ }\href
  {\doibase 10.1103/PhysRevLett.115.222301} {\bibfield  {journal} {\bibinfo
  {journal} {Phys. Rev. Lett.}\ }\textbf {\bibinfo {volume} {115}},\ \bibinfo
  {pages} {222301} (\bibinfo {year} {2015}{\natexlab{a}})},\ \Eprint
  {http://arxiv.org/abs/1505.07812} {arXiv:1505.07812 [nucl-ex]} \BibitemShut
  {NoStop}%
\bibitem [{\citenamefont {Moreland}\ \emph {et~al.}(2015)\citenamefont
  {Moreland}, \citenamefont {Bernhard},\ and\ \citenamefont
  {Bass}}]{Moreland:2014oya}%
  \BibitemOpen
  \bibfield  {author} {\bibinfo {author} {\bibfnamefont {J.~S.}\ \bibnamefont
  {Moreland}}, \bibinfo {author} {\bibfnamefont {J.~E.}\ \bibnamefont
  {Bernhard}}, \ and\ \bibinfo {author} {\bibfnamefont {S.~A.}\ \bibnamefont
  {Bass}},\ }\href {\doibase 10.1103/PhysRevC.92.011901} {\bibfield  {journal}
  {\bibinfo  {journal} {Phys. Rev. C}\ }\textbf {\bibinfo {volume} {92}},\
  \bibinfo {pages} {011901} (\bibinfo {year} {2015})},\ \Eprint
  {http://arxiv.org/abs/1412.4708} {arXiv:1412.4708 [nucl-th]} \BibitemShut
  {NoStop}%
\bibitem [{\citenamefont {Wertepny}\ \emph {et~al.}(2021)\citenamefont
  {Wertepny}, \citenamefont {Noronha-Hostler}, \citenamefont {Sievert},
  \citenamefont {Rao},\ and\ \citenamefont
  {Paladino}}]{Noronha-Hostler:2019ytn}%
  \BibitemOpen
  \bibfield  {author} {\bibinfo {author} {\bibfnamefont {D.}~\bibnamefont
  {Wertepny}}, \bibinfo {author} {\bibfnamefont {J.}~\bibnamefont
  {Noronha-Hostler}}, \bibinfo {author} {\bibfnamefont {M.}~\bibnamefont
  {Sievert}}, \bibinfo {author} {\bibfnamefont {S.}~\bibnamefont {Rao}}, \ and\
  \bibinfo {author} {\bibfnamefont {N.}~\bibnamefont {Paladino}},\ }\href
  {\doibase 10.1016/j.nuclphysa.2020.121839} {\bibfield  {journal} {\bibinfo
  {journal} {Nucl. Phys. A}\ }\textbf {\bibinfo {volume} {1005}},\ \bibinfo
  {pages} {121839} (\bibinfo {year} {2021})},\ \Eprint
  {http://arxiv.org/abs/1905.13323} {arXiv:1905.13323 [hep-ph]} \BibitemShut
  {NoStop}%
\bibitem [{\citenamefont {Nagle}\ and\ \citenamefont
  {Zajc}(2019)}]{Nagle:2018ybc}%
  \BibitemOpen
  \bibfield  {author} {\bibinfo {author} {\bibfnamefont {J.~L.}\ \bibnamefont
  {Nagle}}\ and\ \bibinfo {author} {\bibfnamefont {W.~A.}\ \bibnamefont
  {Zajc}},\ }\href {\doibase 10.1103/PhysRevC.99.054908} {\bibfield  {journal}
  {\bibinfo  {journal} {Phys. Rev. C}\ }\textbf {\bibinfo {volume} {99}},\
  \bibinfo {pages} {054908} (\bibinfo {year} {2019})},\ \Eprint
  {http://arxiv.org/abs/1808.01276} {arXiv:1808.01276 [nucl-th]} \BibitemShut
  {NoStop}%
\bibitem [{\citenamefont {Lappi}(2006)}]{Lappi:2006hq}%
  \BibitemOpen
  \bibfield  {author} {\bibinfo {author} {\bibfnamefont {T.}~\bibnamefont
  {Lappi}},\ }\href {\doibase 10.1016/j.physletb.2006.10.017} {\bibfield
  {journal} {\bibinfo  {journal} {Phys. Lett. B}\ }\textbf {\bibinfo {volume}
  {643}},\ \bibinfo {pages} {11} (\bibinfo {year} {2006})},\ \Eprint
  {http://arxiv.org/abs/hep-ph/0606207} {arXiv:hep-ph/0606207} \BibitemShut
  {NoStop}%
\bibitem [{\citenamefont {Chen}\ \emph {et~al.}(2015)\citenamefont {Chen},
  \citenamefont {Fries}, \citenamefont {Kapusta},\ and\ \citenamefont
  {Li}}]{Chen:2015wia}%
  \BibitemOpen
  \bibfield  {author} {\bibinfo {author} {\bibfnamefont {G.}~\bibnamefont
  {Chen}}, \bibinfo {author} {\bibfnamefont {R.~J.}\ \bibnamefont {Fries}},
  \bibinfo {author} {\bibfnamefont {J.~I.}\ \bibnamefont {Kapusta}}, \ and\
  \bibinfo {author} {\bibfnamefont {Y.}~\bibnamefont {Li}},\ }\href {\doibase
  10.1103/PhysRevC.92.064912} {\bibfield  {journal} {\bibinfo  {journal} {Phys.
  Rev. C}\ }\textbf {\bibinfo {volume} {92}},\ \bibinfo {pages} {064912}
  (\bibinfo {year} {2015})},\ \Eprint {http://arxiv.org/abs/1507.03524}
  {arXiv:1507.03524 [nucl-th]} \BibitemShut {NoStop}%
\bibitem [{\citenamefont {Romatschke}\ and\ \citenamefont
  {Romatschke}(2019)}]{Romatschke:2017ejr}%
  \BibitemOpen
  \bibfield  {author} {\bibinfo {author} {\bibfnamefont {P.}~\bibnamefont
  {Romatschke}}\ and\ \bibinfo {author} {\bibfnamefont {U.}~\bibnamefont
  {Romatschke}},\ }\href {\doibase 10.1017/9781108651998} {\emph {\bibinfo
  {title} {{Relativistic Fluid Dynamics In and Out of Equilibrium}}}},\
  Cambridge Monographs on Mathematical Physics\ (\bibinfo  {publisher}
  {Cambridge University Press},\ \bibinfo {year} {2019})\ \Eprint
  {http://arxiv.org/abs/1712.05815} {arXiv:1712.05815 [nucl-th]} \BibitemShut
  {NoStop}%
\bibitem [{\citenamefont {Renk}\ and\ \citenamefont
  {Niemi}(2014)}]{Renk:2014jja}%
  \BibitemOpen
  \bibfield  {author} {\bibinfo {author} {\bibfnamefont {T.}~\bibnamefont
  {Renk}}\ and\ \bibinfo {author} {\bibfnamefont {H.}~\bibnamefont {Niemi}},\
  }\href {\doibase 10.1103/PhysRevC.89.064907} {\bibfield  {journal} {\bibinfo
  {journal} {Phys. Rev. C}\ }\textbf {\bibinfo {volume} {89}},\ \bibinfo
  {pages} {064907} (\bibinfo {year} {2014})},\ \Eprint
  {http://arxiv.org/abs/1401.2069} {arXiv:1401.2069 [nucl-th]} \BibitemShut
  {NoStop}%
\bibitem [{\citenamefont {Giacalone}\ \emph {et~al.}(2017)\citenamefont
  {Giacalone}, \citenamefont {Noronha-Hostler},\ and\ \citenamefont
  {Ollitrault}}]{Giacalone:2017uqx}%
  \BibitemOpen
  \bibfield  {author} {\bibinfo {author} {\bibfnamefont {G.}~\bibnamefont
  {Giacalone}}, \bibinfo {author} {\bibfnamefont {J.}~\bibnamefont
  {Noronha-Hostler}}, \ and\ \bibinfo {author} {\bibfnamefont {J.-Y.}\
  \bibnamefont {Ollitrault}},\ }\href {\doibase 10.1103/PhysRevC.95.054910}
  {\bibfield  {journal} {\bibinfo  {journal} {Phys. Rev. C}\ }\textbf {\bibinfo
  {volume} {95}},\ \bibinfo {pages} {054910} (\bibinfo {year} {2017})},\
  \Eprint {http://arxiv.org/abs/1702.01730} {arXiv:1702.01730 [nucl-th]}
  \BibitemShut {NoStop}%
\bibitem [{\citenamefont {McLerran}\ and\ \citenamefont
  {Venugopalan}(1994{\natexlab{a}})}]{McLerran:1993ni}%
  \BibitemOpen
  \bibfield  {author} {\bibinfo {author} {\bibfnamefont {L.~D.}\ \bibnamefont
  {McLerran}}\ and\ \bibinfo {author} {\bibfnamefont {R.}~\bibnamefont
  {Venugopalan}},\ }\href {\doibase 10.1103/PhysRevD.49.2233} {\bibfield
  {journal} {\bibinfo  {journal} {Phys. Rev. D}\ }\textbf {\bibinfo {volume}
  {49}},\ \bibinfo {pages} {2233} (\bibinfo {year} {1994}{\natexlab{a}})},\
  \Eprint {http://arxiv.org/abs/hep-ph/9309289} {arXiv:hep-ph/9309289}
  \BibitemShut {NoStop}%
\bibitem [{\citenamefont {McLerran}\ and\ \citenamefont
  {Venugopalan}(1994{\natexlab{b}})}]{McLerran:1993ka}%
  \BibitemOpen
  \bibfield  {author} {\bibinfo {author} {\bibfnamefont {L.~D.}\ \bibnamefont
  {McLerran}}\ and\ \bibinfo {author} {\bibfnamefont {R.}~\bibnamefont
  {Venugopalan}},\ }\href {\doibase 10.1103/PhysRevD.49.3352} {\bibfield
  {journal} {\bibinfo  {journal} {Phys. Rev. D}\ }\textbf {\bibinfo {volume}
  {49}},\ \bibinfo {pages} {3352} (\bibinfo {year} {1994}{\natexlab{b}})},\
  \Eprint {http://arxiv.org/abs/hep-ph/9311205} {arXiv:hep-ph/9311205}
  \BibitemShut {NoStop}%
\bibitem [{\citenamefont {McLerran}\ and\ \citenamefont
  {Venugopalan}(1994{\natexlab{c}})}]{McLerran:1994vd}%
  \BibitemOpen
  \bibfield  {author} {\bibinfo {author} {\bibfnamefont {L.~D.}\ \bibnamefont
  {McLerran}}\ and\ \bibinfo {author} {\bibfnamefont {R.}~\bibnamefont
  {Venugopalan}},\ }\href {\doibase 10.1103/PhysRevD.50.2225} {\bibfield
  {journal} {\bibinfo  {journal} {Phys. Rev. D}\ }\textbf {\bibinfo {volume}
  {50}},\ \bibinfo {pages} {2225} (\bibinfo {year} {1994}{\natexlab{c}})},\
  \Eprint {http://arxiv.org/abs/hep-ph/9402335} {arXiv:hep-ph/9402335}
  \BibitemShut {NoStop}%
\bibitem [{\citenamefont {Schenke}\ \emph {et~al.}(2012)\citenamefont
  {Schenke}, \citenamefont {Tribedy},\ and\ \citenamefont
  {Venugopalan}}]{Schenke:2012wb}%
  \BibitemOpen
  \bibfield  {author} {\bibinfo {author} {\bibfnamefont {B.}~\bibnamefont
  {Schenke}}, \bibinfo {author} {\bibfnamefont {P.}~\bibnamefont {Tribedy}}, \
  and\ \bibinfo {author} {\bibfnamefont {R.}~\bibnamefont {Venugopalan}},\
  }\href {\doibase 10.1103/PhysRevLett.108.252301} {\bibfield  {journal}
  {\bibinfo  {journal} {Phys. Rev. Lett.}\ }\textbf {\bibinfo {volume} {108}},\
  \bibinfo {pages} {252301} (\bibinfo {year} {2012})},\ \Eprint
  {http://arxiv.org/abs/1202.6646} {arXiv:1202.6646 [nucl-th]} \BibitemShut
  {NoStop}%
\bibitem [{\citenamefont {McLerran}\ and\ \citenamefont
  {Tribedy}(2016)}]{McLerran:2015qxa}%
  \BibitemOpen
  \bibfield  {author} {\bibinfo {author} {\bibfnamefont {L.}~\bibnamefont
  {McLerran}}\ and\ \bibinfo {author} {\bibfnamefont {P.}~\bibnamefont
  {Tribedy}},\ }\href {\doibase 10.1016/j.nuclphysa.2015.10.008} {\bibfield
  {journal} {\bibinfo  {journal} {Nucl. Phys. A}\ }\textbf {\bibinfo {volume}
  {945}},\ \bibinfo {pages} {216} (\bibinfo {year} {2016})},\ \Eprint
  {http://arxiv.org/abs/1508.03292} {arXiv:1508.03292 [hep-ph]} \BibitemShut
  {NoStop}%
\bibitem [{\citenamefont {Mace}\ \emph {et~al.}(2018)\citenamefont {Mace},
  \citenamefont {Skokov}, \citenamefont {Tribedy},\ and\ \citenamefont
  {Venugopalan}}]{Mace:2018vwq}%
  \BibitemOpen
  \bibfield  {author} {\bibinfo {author} {\bibfnamefont {M.}~\bibnamefont
  {Mace}}, \bibinfo {author} {\bibfnamefont {V.~V.}\ \bibnamefont {Skokov}},
  \bibinfo {author} {\bibfnamefont {P.}~\bibnamefont {Tribedy}}, \ and\
  \bibinfo {author} {\bibfnamefont {R.}~\bibnamefont {Venugopalan}},\ }\href
  {\doibase 10.1103/PhysRevLett.121.052301} {\bibfield  {journal} {\bibinfo
  {journal} {Phys. Rev. Lett.}\ }\textbf {\bibinfo {volume} {121}},\ \bibinfo
  {pages} {052301} (\bibinfo {year} {2018})},\ \bibinfo {note} {[Erratum:
  Phys.Rev.Lett. 123, 039901 (2019)]},\ \Eprint
  {http://arxiv.org/abs/1805.09342} {arXiv:1805.09342 [hep-ph]} \BibitemShut
  {NoStop}%
\bibitem [{\citenamefont {Moreland}\ \emph {et~al.}(2020)\citenamefont
  {Moreland}, \citenamefont {Bernhard},\ and\ \citenamefont
  {Bass}}]{Moreland:2018gsh}%
  \BibitemOpen
  \bibfield  {author} {\bibinfo {author} {\bibfnamefont {J.~S.}\ \bibnamefont
  {Moreland}}, \bibinfo {author} {\bibfnamefont {J.~E.}\ \bibnamefont
  {Bernhard}}, \ and\ \bibinfo {author} {\bibfnamefont {S.~A.}\ \bibnamefont
  {Bass}},\ }\href {\doibase 10.1103/PhysRevC.101.024911} {\bibfield  {journal}
  {\bibinfo  {journal} {Phys. Rev. C}\ }\textbf {\bibinfo {volume} {101}},\
  \bibinfo {pages} {024911} (\bibinfo {year} {2020})},\ \Eprint
  {http://arxiv.org/abs/1808.02106} {arXiv:1808.02106 [nucl-th]} \BibitemShut
  {NoStop}%
\bibitem [{\citenamefont {Everett}\ \emph {et~al.}(2021)\citenamefont {Everett}
  \emph {et~al.}}]{Everett:2020xug}%
  \BibitemOpen
  \bibfield  {author} {\bibinfo {author} {\bibfnamefont {D.}~\bibnamefont
  {Everett}} \emph {et~al.} (\bibinfo {collaboration} {JETSCAPE}),\ }\href
  {\doibase 10.1103/PhysRevC.103.054904} {\bibfield  {journal} {\bibinfo
  {journal} {Phys. Rev. C}\ }\textbf {\bibinfo {volume} {103}},\ \bibinfo
  {pages} {054904} (\bibinfo {year} {2021})},\ \Eprint
  {http://arxiv.org/abs/2011.01430} {arXiv:2011.01430 [hep-ph]} \BibitemShut
  {NoStop}%
\bibitem [{\citenamefont {Nijs}\ \emph {et~al.}(2021)\citenamefont {Nijs},
  \citenamefont {Van Der~Schee}, \citenamefont {G\"ursoy},\ and\ \citenamefont
  {Snellings}}]{Nijs:2020roc}%
  \BibitemOpen
  \bibfield  {author} {\bibinfo {author} {\bibfnamefont {G.}~\bibnamefont
  {Nijs}}, \bibinfo {author} {\bibfnamefont {W.}~\bibnamefont {Van Der~Schee}},
  \bibinfo {author} {\bibfnamefont {U.}~\bibnamefont {G\"ursoy}}, \ and\
  \bibinfo {author} {\bibfnamefont {R.}~\bibnamefont {Snellings}},\ }\href
  {\doibase 10.1103/PhysRevC.103.054909} {\bibfield  {journal} {\bibinfo
  {journal} {Phys. Rev. C}\ }\textbf {\bibinfo {volume} {103}},\ \bibinfo
  {pages} {054909} (\bibinfo {year} {2021})},\ \Eprint
  {http://arxiv.org/abs/2010.15134} {arXiv:2010.15134 [nucl-th]} \BibitemShut
  {NoStop}%
\bibitem [{\citenamefont {Chatrchyan}\ \emph {et~al.}(2013)\citenamefont
  {Chatrchyan} \emph {et~al.}}]{Chatrchyan:2013nka}%
  \BibitemOpen
  \bibfield  {author} {\bibinfo {author} {\bibfnamefont {S.}~\bibnamefont
  {Chatrchyan}} \emph {et~al.} (\bibinfo {collaboration} {CMS}),\ }\href
  {\doibase 10.1016/j.physletb.2013.06.028} {\bibfield  {journal} {\bibinfo
  {journal} {Phys. Lett. B}\ }\textbf {\bibinfo {volume} {724}},\ \bibinfo
  {pages} {213} (\bibinfo {year} {2013})},\ \Eprint
  {http://arxiv.org/abs/1305.0609} {arXiv:1305.0609 [nucl-ex]} \BibitemShut
  {NoStop}%
\bibitem [{\citenamefont {Aaboud}\ \emph {et~al.}(2017)\citenamefont {Aaboud}
  \emph {et~al.}}]{Aaboud:2017acw}%
  \BibitemOpen
  \bibfield  {author} {\bibinfo {author} {\bibfnamefont {M.}~\bibnamefont
  {Aaboud}} \emph {et~al.} (\bibinfo {collaboration} {ATLAS}),\ }\href
  {\doibase 10.1140/epjc/s10052-017-4988-1} {\bibfield  {journal} {\bibinfo
  {journal} {Eur. Phys. J. C}\ }\textbf {\bibinfo {volume} {77}},\ \bibinfo
  {pages} {428} (\bibinfo {year} {2017})},\ \Eprint
  {http://arxiv.org/abs/1705.04176} {arXiv:1705.04176 [hep-ex]} \BibitemShut
  {NoStop}%
\bibitem [{\citenamefont {Aaboud}\ \emph {et~al.}(2018)\citenamefont {Aaboud}
  \emph {et~al.}}]{Aaboud:2017blb}%
  \BibitemOpen
  \bibfield  {author} {\bibinfo {author} {\bibfnamefont {M.}~\bibnamefont
  {Aaboud}} \emph {et~al.} (\bibinfo {collaboration} {ATLAS}),\ }\href
  {\doibase 10.1103/PhysRevC.97.024904} {\bibfield  {journal} {\bibinfo
  {journal} {Phys. Rev. C}\ }\textbf {\bibinfo {volume} {97}},\ \bibinfo
  {pages} {024904} (\bibinfo {year} {2018})},\ \Eprint
  {http://arxiv.org/abs/1708.03559} {arXiv:1708.03559 [hep-ex]} \BibitemShut
  {NoStop}%
\bibitem [{\citenamefont {Aad}\ \emph {et~al.}(2013)\citenamefont {Aad} \emph
  {et~al.}}]{Aad:2013fja}%
  \BibitemOpen
  \bibfield  {author} {\bibinfo {author} {\bibfnamefont {G.}~\bibnamefont
  {Aad}} \emph {et~al.} (\bibinfo {collaboration} {ATLAS}),\ }\href {\doibase
  10.1016/j.physletb.2013.06.057} {\bibfield  {journal} {\bibinfo  {journal}
  {Phys. Lett. B}\ }\textbf {\bibinfo {volume} {725}},\ \bibinfo {pages} {60}
  (\bibinfo {year} {2013})},\ \Eprint {http://arxiv.org/abs/1303.2084}
  {arXiv:1303.2084 [hep-ex]} \BibitemShut {NoStop}%
\bibitem [{\citenamefont {Sirunyan}\ \emph
  {et~al.}(2018{\natexlab{a}})\citenamefont {Sirunyan} \emph
  {et~al.}}]{Sirunyan:2018toe}%
  \BibitemOpen
  \bibfield  {author} {\bibinfo {author} {\bibfnamefont {A.~M.}\ \bibnamefont
  {Sirunyan}} \emph {et~al.} (\bibinfo {collaboration} {CMS}),\ }\href
  {\doibase 10.1103/PhysRevLett.121.082301} {\bibfield  {journal} {\bibinfo
  {journal} {Phys. Rev. Lett.}\ }\textbf {\bibinfo {volume} {121}},\ \bibinfo
  {pages} {082301} (\bibinfo {year} {2018}{\natexlab{a}})},\ \Eprint
  {http://arxiv.org/abs/1804.09767} {arXiv:1804.09767 [hep-ex]} \BibitemShut
  {NoStop}%
\bibitem [{\citenamefont {Khachatryan}\ \emph
  {et~al.}(2015{\natexlab{a}})\citenamefont {Khachatryan} \emph
  {et~al.}}]{Khachatryan:2014jra}%
  \BibitemOpen
  \bibfield  {author} {\bibinfo {author} {\bibfnamefont {V.}~\bibnamefont
  {Khachatryan}} \emph {et~al.} (\bibinfo {collaboration} {CMS}),\ }\href
  {\doibase 10.1016/j.physletb.2015.01.034} {\bibfield  {journal} {\bibinfo
  {journal} {Phys. Lett. B}\ }\textbf {\bibinfo {volume} {742}},\ \bibinfo
  {pages} {200} (\bibinfo {year} {2015}{\natexlab{a}})},\ \Eprint
  {http://arxiv.org/abs/1409.3392} {arXiv:1409.3392 [nucl-ex]} \BibitemShut
  {NoStop}%
\bibitem [{\citenamefont {Khachatryan}\ \emph
  {et~al.}(2015{\natexlab{b}})\citenamefont {Khachatryan} \emph
  {et~al.}}]{Khachatryan:2015waa}%
  \BibitemOpen
  \bibfield  {author} {\bibinfo {author} {\bibfnamefont {V.}~\bibnamefont
  {Khachatryan}} \emph {et~al.} (\bibinfo {collaboration} {CMS}),\ }\href
  {\doibase 10.1103/PhysRevLett.115.012301} {\bibfield  {journal} {\bibinfo
  {journal} {Phys. Rev. Lett.}\ }\textbf {\bibinfo {volume} {115}},\ \bibinfo
  {pages} {012301} (\bibinfo {year} {2015}{\natexlab{b}})},\ \Eprint
  {http://arxiv.org/abs/1502.05382} {arXiv:1502.05382 [nucl-ex]} \BibitemShut
  {NoStop}%
\bibitem [{\citenamefont {Khachatryan}\ \emph
  {et~al.}(2015{\natexlab{c}})\citenamefont {Khachatryan} \emph
  {et~al.}}]{Khachatryan:2015oea}%
  \BibitemOpen
  \bibfield  {author} {\bibinfo {author} {\bibfnamefont {V.}~\bibnamefont
  {Khachatryan}} \emph {et~al.} (\bibinfo {collaboration} {CMS}),\ }\href
  {\doibase 10.1103/PhysRevC.92.034911} {\bibfield  {journal} {\bibinfo
  {journal} {Phys. Rev. C}\ }\textbf {\bibinfo {volume} {92}},\ \bibinfo
  {pages} {034911} (\bibinfo {year} {2015}{\natexlab{c}})},\ \Eprint
  {http://arxiv.org/abs/1503.01692} {arXiv:1503.01692 [nucl-ex]} \BibitemShut
  {NoStop}%
\bibitem [{\citenamefont {Sirunyan}\ \emph
  {et~al.}(2018{\natexlab{b}})\citenamefont {Sirunyan} \emph
  {et~al.}}]{Sirunyan:2017uyl}%
  \BibitemOpen
  \bibfield  {author} {\bibinfo {author} {\bibfnamefont {A.~M.}\ \bibnamefont
  {Sirunyan}} \emph {et~al.} (\bibinfo {collaboration} {CMS}),\ }\href
  {\doibase 10.1103/PhysRevLett.120.092301} {\bibfield  {journal} {\bibinfo
  {journal} {Phys. Rev. Lett.}\ }\textbf {\bibinfo {volume} {120}},\ \bibinfo
  {pages} {092301} (\bibinfo {year} {2018}{\natexlab{b}})},\ \Eprint
  {http://arxiv.org/abs/1709.09189} {arXiv:1709.09189 [nucl-ex]} \BibitemShut
  {NoStop}%
\bibitem [{\citenamefont {Abelev}\ \emph {et~al.}(2013)\citenamefont {Abelev}
  \emph {et~al.}}]{ABELEV:2013wsa}%
  \BibitemOpen
  \bibfield  {author} {\bibinfo {author} {\bibfnamefont {B.~B.}\ \bibnamefont
  {Abelev}} \emph {et~al.} (\bibinfo {collaboration} {ALICE}),\ }\href
  {\doibase 10.1016/j.physletb.2013.08.024} {\bibfield  {journal} {\bibinfo
  {journal} {Phys. Lett. B}\ }\textbf {\bibinfo {volume} {726}},\ \bibinfo
  {pages} {164} (\bibinfo {year} {2013})},\ \Eprint
  {http://arxiv.org/abs/1307.3237} {arXiv:1307.3237 [nucl-ex]} \BibitemShut
  {NoStop}%
\bibitem [{\citenamefont {Abelev}\ \emph {et~al.}(2014)\citenamefont {Abelev}
  \emph {et~al.}}]{Abelev:2014mda}%
  \BibitemOpen
  \bibfield  {author} {\bibinfo {author} {\bibfnamefont {B.~B.}\ \bibnamefont
  {Abelev}} \emph {et~al.} (\bibinfo {collaboration} {ALICE}),\ }\href
  {\doibase 10.1103/PhysRevC.90.054901} {\bibfield  {journal} {\bibinfo
  {journal} {Phys. Rev. C}\ }\textbf {\bibinfo {volume} {90}},\ \bibinfo
  {pages} {054901} (\bibinfo {year} {2014})},\ \Eprint
  {http://arxiv.org/abs/1406.2474} {arXiv:1406.2474 [nucl-ex]} \BibitemShut
  {NoStop}%
\bibitem [{\citenamefont {Bozek}(2012)}]{Bozek:2011if}%
  \BibitemOpen
  \bibfield  {author} {\bibinfo {author} {\bibfnamefont {P.}~\bibnamefont
  {Bozek}},\ }\href {\doibase 10.1103/PhysRevC.85.014911} {\bibfield  {journal}
  {\bibinfo  {journal} {Phys. Rev. C}\ }\textbf {\bibinfo {volume} {85}},\
  \bibinfo {pages} {014911} (\bibinfo {year} {2012})},\ \Eprint
  {http://arxiv.org/abs/1112.0915} {arXiv:1112.0915 [hep-ph]} \BibitemShut
  {NoStop}%
\bibitem [{\citenamefont {Bozek}\ and\ \citenamefont
  {Broniowski}(2013{\natexlab{b}})}]{Bozek:2012gr}%
  \BibitemOpen
  \bibfield  {author} {\bibinfo {author} {\bibfnamefont {P.}~\bibnamefont
  {Bozek}}\ and\ \bibinfo {author} {\bibfnamefont {W.}~\bibnamefont
  {Broniowski}},\ }\href {\doibase 10.1016/j.physletb.2012.12.051} {\bibfield
  {journal} {\bibinfo  {journal} {Phys. Lett. B}\ }\textbf {\bibinfo {volume}
  {718}},\ \bibinfo {pages} {1557} (\bibinfo {year} {2013}{\natexlab{b}})},\
  \Eprint {http://arxiv.org/abs/1211.0845} {arXiv:1211.0845 [nucl-th]}
  \BibitemShut {NoStop}%
\bibitem [{\citenamefont {Bozek}\ \emph {et~al.}(2013)\citenamefont {Bozek},
  \citenamefont {Broniowski},\ and\ \citenamefont {Torrieri}}]{Bozek:2013ska}%
  \BibitemOpen
  \bibfield  {author} {\bibinfo {author} {\bibfnamefont {P.}~\bibnamefont
  {Bozek}}, \bibinfo {author} {\bibfnamefont {W.}~\bibnamefont {Broniowski}}, \
  and\ \bibinfo {author} {\bibfnamefont {G.}~\bibnamefont {Torrieri}},\ }\href
  {\doibase 10.1103/PhysRevLett.111.172303} {\bibfield  {journal} {\bibinfo
  {journal} {Phys. Rev. Lett.}\ }\textbf {\bibinfo {volume} {111}},\ \bibinfo
  {pages} {172303} (\bibinfo {year} {2013})},\ \Eprint
  {http://arxiv.org/abs/1307.5060} {arXiv:1307.5060 [nucl-th]} \BibitemShut
  {NoStop}%
\bibitem [{\citenamefont {Kozlov}\ \emph {et~al.}(2014)\citenamefont {Kozlov},
  \citenamefont {Luzum}, \citenamefont {Denicol}, \citenamefont {Jeon},\ and\
  \citenamefont {Gale}}]{Kozlov:2014fqa}%
  \BibitemOpen
  \bibfield  {author} {\bibinfo {author} {\bibfnamefont {I.}~\bibnamefont
  {Kozlov}}, \bibinfo {author} {\bibfnamefont {M.}~\bibnamefont {Luzum}},
  \bibinfo {author} {\bibfnamefont {G.}~\bibnamefont {Denicol}}, \bibinfo
  {author} {\bibfnamefont {S.}~\bibnamefont {Jeon}}, \ and\ \bibinfo {author}
  {\bibfnamefont {C.}~\bibnamefont {Gale}},\ }\href@noop {} {\  (\bibinfo
  {year} {2014})},\ \Eprint {http://arxiv.org/abs/1405.3976} {arXiv:1405.3976
  [nucl-th]} \BibitemShut {NoStop}%
\bibitem [{\citenamefont {Zhou}\ \emph {et~al.}(2015)\citenamefont {Zhou},
  \citenamefont {Zhu}, \citenamefont {Li},\ and\ \citenamefont
  {Song}}]{Zhou:2015iba}%
  \BibitemOpen
  \bibfield  {author} {\bibinfo {author} {\bibfnamefont {Y.}~\bibnamefont
  {Zhou}}, \bibinfo {author} {\bibfnamefont {X.}~\bibnamefont {Zhu}}, \bibinfo
  {author} {\bibfnamefont {P.}~\bibnamefont {Li}}, \ and\ \bibinfo {author}
  {\bibfnamefont {H.}~\bibnamefont {Song}},\ }\href {\doibase
  10.1103/PhysRevC.91.064908} {\bibfield  {journal} {\bibinfo  {journal} {Phys.
  Rev. C}\ }\textbf {\bibinfo {volume} {91}},\ \bibinfo {pages} {064908}
  (\bibinfo {year} {2015})},\ \Eprint {http://arxiv.org/abs/1503.06986}
  {arXiv:1503.06986 [nucl-th]} \BibitemShut {NoStop}%
\bibitem [{\citenamefont {Zhao}\ \emph {et~al.}(2018)\citenamefont {Zhao},
  \citenamefont {Zhou}, \citenamefont {Xu}, \citenamefont {Deng},\ and\
  \citenamefont {Song}}]{Zhao:2017rgg}%
  \BibitemOpen
  \bibfield  {author} {\bibinfo {author} {\bibfnamefont {W.}~\bibnamefont
  {Zhao}}, \bibinfo {author} {\bibfnamefont {Y.}~\bibnamefont {Zhou}}, \bibinfo
  {author} {\bibfnamefont {H.}~\bibnamefont {Xu}}, \bibinfo {author}
  {\bibfnamefont {W.}~\bibnamefont {Deng}}, \ and\ \bibinfo {author}
  {\bibfnamefont {H.}~\bibnamefont {Song}},\ }\href {\doibase
  10.1016/j.physletb.2018.03.022} {\bibfield  {journal} {\bibinfo  {journal}
  {Phys. Lett. B}\ }\textbf {\bibinfo {volume} {780}},\ \bibinfo {pages} {495}
  (\bibinfo {year} {2018})},\ \Eprint {http://arxiv.org/abs/1801.00271}
  {arXiv:1801.00271 [nucl-th]} \BibitemShut {NoStop}%
\bibitem [{\citenamefont {M\"antysaari}\ \emph {et~al.}(2017)\citenamefont
  {M\"antysaari}, \citenamefont {Schenke}, \citenamefont {Shen},\ and\
  \citenamefont {Tribedy}}]{Mantysaari:2017cni}%
  \BibitemOpen
  \bibfield  {author} {\bibinfo {author} {\bibfnamefont {H.}~\bibnamefont
  {M\"antysaari}}, \bibinfo {author} {\bibfnamefont {B.}~\bibnamefont
  {Schenke}}, \bibinfo {author} {\bibfnamefont {C.}~\bibnamefont {Shen}}, \
  and\ \bibinfo {author} {\bibfnamefont {P.}~\bibnamefont {Tribedy}},\ }\href
  {\doibase 10.1016/j.physletb.2017.07.038} {\bibfield  {journal} {\bibinfo
  {journal} {Phys. Lett. B}\ }\textbf {\bibinfo {volume} {772}},\ \bibinfo
  {pages} {681} (\bibinfo {year} {2017})},\ \Eprint
  {http://arxiv.org/abs/1705.03177} {arXiv:1705.03177 [nucl-th]} \BibitemShut
  {NoStop}%
\bibitem [{\citenamefont {Greif}\ \emph {et~al.}(2017)\citenamefont {Greif},
  \citenamefont {Greiner}, \citenamefont {Schenke}, \citenamefont
  {Schlichting},\ and\ \citenamefont {Xu}}]{Greif:2017bnr}%
  \BibitemOpen
  \bibfield  {author} {\bibinfo {author} {\bibfnamefont {M.}~\bibnamefont
  {Greif}}, \bibinfo {author} {\bibfnamefont {C.}~\bibnamefont {Greiner}},
  \bibinfo {author} {\bibfnamefont {B.}~\bibnamefont {Schenke}}, \bibinfo
  {author} {\bibfnamefont {S.}~\bibnamefont {Schlichting}}, \ and\ \bibinfo
  {author} {\bibfnamefont {Z.}~\bibnamefont {Xu}},\ }\href {\doibase
  10.1103/PhysRevD.96.091504} {\bibfield  {journal} {\bibinfo  {journal} {Phys.
  Rev. D}\ }\textbf {\bibinfo {volume} {96}},\ \bibinfo {pages} {091504}
  (\bibinfo {year} {2017})},\ \Eprint {http://arxiv.org/abs/1708.02076}
  {arXiv:1708.02076 [hep-ph]} \BibitemShut {NoStop}%
\bibitem [{\citenamefont {Schenke}\ \emph {et~al.}(2016)\citenamefont
  {Schenke}, \citenamefont {Schlichting}, \citenamefont {Tribedy},\ and\
  \citenamefont {Venugopalan}}]{Schenke:2016lrs}%
  \BibitemOpen
  \bibfield  {author} {\bibinfo {author} {\bibfnamefont {B.}~\bibnamefont
  {Schenke}}, \bibinfo {author} {\bibfnamefont {S.}~\bibnamefont
  {Schlichting}}, \bibinfo {author} {\bibfnamefont {P.}~\bibnamefont
  {Tribedy}}, \ and\ \bibinfo {author} {\bibfnamefont {R.}~\bibnamefont
  {Venugopalan}},\ }\href {\doibase 10.1103/PhysRevLett.117.162301} {\bibfield
  {journal} {\bibinfo  {journal} {Phys. Rev. Lett.}\ }\textbf {\bibinfo
  {volume} {117}},\ \bibinfo {pages} {162301} (\bibinfo {year} {2016})},\
  \Eprint {http://arxiv.org/abs/1607.02496} {arXiv:1607.02496 [hep-ph]}
  \BibitemShut {NoStop}%
\bibitem [{\citenamefont {M\"antysaari}\ and\ \citenamefont
  {Schenke}(2016)}]{Mantysaari:2016ykx}%
  \BibitemOpen
  \bibfield  {author} {\bibinfo {author} {\bibfnamefont {H.}~\bibnamefont
  {M\"antysaari}}\ and\ \bibinfo {author} {\bibfnamefont {B.}~\bibnamefont
  {Schenke}},\ }\href {\doibase 10.1103/PhysRevLett.117.052301} {\bibfield
  {journal} {\bibinfo  {journal} {Phys. Rev. Lett.}\ }\textbf {\bibinfo
  {volume} {117}},\ \bibinfo {pages} {052301} (\bibinfo {year} {2016})},\
  \Eprint {http://arxiv.org/abs/1603.04349} {arXiv:1603.04349 [hep-ph]}
  \BibitemShut {NoStop}%
\bibitem [{\citenamefont {Albacete}\ \emph {et~al.}(2018)\citenamefont
  {Albacete}, \citenamefont {Petersen},\ and\ \citenamefont
  {Soto-Ontoso}}]{Albacete:2017ajt}%
  \BibitemOpen
  \bibfield  {author} {\bibinfo {author} {\bibfnamefont {J.~L.}\ \bibnamefont
  {Albacete}}, \bibinfo {author} {\bibfnamefont {H.}~\bibnamefont {Petersen}},
  \ and\ \bibinfo {author} {\bibfnamefont {A.}~\bibnamefont {Soto-Ontoso}},\
  }\href {\doibase 10.1016/j.physletb.2018.01.011} {\bibfield  {journal}
  {\bibinfo  {journal} {Phys. Lett. B}\ }\textbf {\bibinfo {volume} {778}},\
  \bibinfo {pages} {128} (\bibinfo {year} {2018})},\ \Eprint
  {http://arxiv.org/abs/1707.05592} {arXiv:1707.05592 [hep-ph]} \BibitemShut
  {NoStop}%
\bibitem [{\citenamefont {Aidala}\ \emph {et~al.}(2019)\citenamefont {Aidala}
  \emph {et~al.}}]{Aidala:2018mcw}%
  \BibitemOpen
  \bibfield  {author} {\bibinfo {author} {\bibfnamefont {C.}~\bibnamefont
  {Aidala}} \emph {et~al.} (\bibinfo {collaboration} {PHENIX}),\ }\href
  {\doibase 10.1038/s41567-018-0360-0} {\bibfield  {journal} {\bibinfo
  {journal} {Nature Phys.}\ }\textbf {\bibinfo {volume} {15}},\ \bibinfo
  {pages} {214} (\bibinfo {year} {2019})},\ \Eprint
  {http://arxiv.org/abs/1805.02973} {arXiv:1805.02973 [nucl-ex]} \BibitemShut
  {NoStop}%
\bibitem [{\citenamefont {Adare}\ \emph {et~al.}(2018)\citenamefont {Adare}
  \emph {et~al.}}]{Adare:2018toe}%
  \BibitemOpen
  \bibfield  {author} {\bibinfo {author} {\bibfnamefont {A.}~\bibnamefont
  {Adare}} \emph {et~al.} (\bibinfo {collaboration} {PHENIX}),\ }\href
  {\doibase 10.1103/PhysRevLett.121.222301} {\bibfield  {journal} {\bibinfo
  {journal} {Phys. Rev. Lett.}\ }\textbf {\bibinfo {volume} {121}},\ \bibinfo
  {pages} {222301} (\bibinfo {year} {2018})},\ \Eprint
  {http://arxiv.org/abs/1807.11928} {arXiv:1807.11928 [nucl-ex]} \BibitemShut
  {NoStop}%
\bibitem [{\citenamefont {Schenke}\ \emph {et~al.}(2020)\citenamefont
  {Schenke}, \citenamefont {Shen},\ and\ \citenamefont
  {Tribedy}}]{Schenke:2019pmk}%
  \BibitemOpen
  \bibfield  {author} {\bibinfo {author} {\bibfnamefont {B.}~\bibnamefont
  {Schenke}}, \bibinfo {author} {\bibfnamefont {C.}~\bibnamefont {Shen}}, \
  and\ \bibinfo {author} {\bibfnamefont {P.}~\bibnamefont {Tribedy}},\ }\href
  {\doibase 10.1016/j.physletb.2020.135322} {\bibfield  {journal} {\bibinfo
  {journal} {Phys. Lett. B}\ }\textbf {\bibinfo {volume} {803}},\ \bibinfo
  {pages} {135322} (\bibinfo {year} {2020})},\ \Eprint
  {http://arxiv.org/abs/1908.06212} {arXiv:1908.06212 [nucl-th]} \BibitemShut
  {NoStop}%
\bibitem [{\citenamefont {Bilandzic}\ \emph {et~al.}(2011)\citenamefont
  {Bilandzic}, \citenamefont {Snellings},\ and\ \citenamefont
  {Voloshin}}]{Bilandzic:2010jr}%
  \BibitemOpen
  \bibfield  {author} {\bibinfo {author} {\bibfnamefont {A.}~\bibnamefont
  {Bilandzic}}, \bibinfo {author} {\bibfnamefont {R.}~\bibnamefont
  {Snellings}}, \ and\ \bibinfo {author} {\bibfnamefont {S.}~\bibnamefont
  {Voloshin}},\ }\href {\doibase 10.1103/PhysRevC.83.044913} {\bibfield
  {journal} {\bibinfo  {journal} {Phys. Rev. C}\ }\textbf {\bibinfo {volume}
  {83}},\ \bibinfo {pages} {044913} (\bibinfo {year} {2011})},\ \Eprint
  {http://arxiv.org/abs/1010.0233} {arXiv:1010.0233 [nucl-ex]} \BibitemShut
  {NoStop}%
\bibitem [{\citenamefont {Abelev}\ \emph {et~al.}(2009)\citenamefont {Abelev}
  \emph {et~al.}}]{Abelev:2008ab}%
  \BibitemOpen
  \bibfield  {author} {\bibinfo {author} {\bibfnamefont {B.~I.}\ \bibnamefont
  {Abelev}} \emph {et~al.} (\bibinfo {collaboration} {STAR}),\ }\href {\doibase
  10.1103/PhysRevC.79.034909} {\bibfield  {journal} {\bibinfo  {journal} {Phys.
  Rev. C}\ }\textbf {\bibinfo {volume} {79}},\ \bibinfo {pages} {034909}
  (\bibinfo {year} {2009})},\ \Eprint {http://arxiv.org/abs/0808.2041}
  {arXiv:0808.2041 [nucl-ex]} \BibitemShut {NoStop}%
\bibitem [{\citenamefont {Rao}\ \emph {et~al.}(2021)\citenamefont {Rao},
  \citenamefont {Sievert},\ and\ \citenamefont
  {Noronha-Hostler}}]{Rao:2019vgy}%
  \BibitemOpen
  \bibfield  {author} {\bibinfo {author} {\bibfnamefont {S.}~\bibnamefont
  {Rao}}, \bibinfo {author} {\bibfnamefont {M.}~\bibnamefont {Sievert}}, \ and\
  \bibinfo {author} {\bibfnamefont {J.}~\bibnamefont {Noronha-Hostler}},\
  }\href {\doibase 10.1103/PhysRevC.103.034910} {\bibfield  {journal} {\bibinfo
   {journal} {Phys. Rev. C}\ }\textbf {\bibinfo {volume} {103}},\ \bibinfo
  {pages} {034910} (\bibinfo {year} {2021})},\ \Eprint
  {http://arxiv.org/abs/1910.03677} {arXiv:1910.03677 [nucl-th]} \BibitemShut
  {NoStop}%
\bibitem [{\citenamefont {Noronha-Hostler}\ \emph {et~al.}(2014)\citenamefont
  {Noronha-Hostler}, \citenamefont {Noronha},\ and\ \citenamefont
  {Grassi}}]{Noronha-Hostler:2014dqa}%
  \BibitemOpen
  \bibfield  {author} {\bibinfo {author} {\bibfnamefont {J.}~\bibnamefont
  {Noronha-Hostler}}, \bibinfo {author} {\bibfnamefont {J.}~\bibnamefont
  {Noronha}}, \ and\ \bibinfo {author} {\bibfnamefont {F.}~\bibnamefont
  {Grassi}},\ }\href {\doibase 10.1103/PhysRevC.90.034907} {\bibfield
  {journal} {\bibinfo  {journal} {Phys. Rev. C}\ }\textbf {\bibinfo {volume}
  {90}},\ \bibinfo {pages} {034907} (\bibinfo {year} {2014})},\ \Eprint
  {http://arxiv.org/abs/1406.3333} {arXiv:1406.3333 [nucl-th]} \BibitemShut
  {NoStop}%
\bibitem [{\citenamefont {Noronha-Hostler}\ \emph {et~al.}(2013)\citenamefont
  {Noronha-Hostler}, \citenamefont {Denicol}, \citenamefont {Noronha},
  \citenamefont {Andrade},\ and\ \citenamefont
  {Grassi}}]{Noronha-Hostler:2013gga}%
  \BibitemOpen
  \bibfield  {author} {\bibinfo {author} {\bibfnamefont {J.}~\bibnamefont
  {Noronha-Hostler}}, \bibinfo {author} {\bibfnamefont {G.~S.}\ \bibnamefont
  {Denicol}}, \bibinfo {author} {\bibfnamefont {J.}~\bibnamefont {Noronha}},
  \bibinfo {author} {\bibfnamefont {R.~P.~G.}\ \bibnamefont {Andrade}}, \ and\
  \bibinfo {author} {\bibfnamefont {F.}~\bibnamefont {Grassi}},\ }\href
  {\doibase 10.1103/PhysRevC.88.044916} {\bibfield  {journal} {\bibinfo
  {journal} {Phys. Rev. C}\ }\textbf {\bibinfo {volume} {88}},\ \bibinfo
  {pages} {044916} (\bibinfo {year} {2013})},\ \Eprint
  {http://arxiv.org/abs/1305.1981} {arXiv:1305.1981 [nucl-th]} \BibitemShut
  {NoStop}%
\bibitem [{\citenamefont {Alba}\ \emph {et~al.}(2017)\citenamefont {Alba} \emph
  {et~al.}}]{Alba:2017mqu}%
  \BibitemOpen
  \bibfield  {author} {\bibinfo {author} {\bibfnamefont {P.}~\bibnamefont
  {Alba}} \emph {et~al.},\ }\href {\doibase 10.1103/PhysRevD.96.034517}
  {\bibfield  {journal} {\bibinfo  {journal} {Phys. Rev. D}\ }\textbf {\bibinfo
  {volume} {96}},\ \bibinfo {pages} {034517} (\bibinfo {year} {2017})},\
  \Eprint {http://arxiv.org/abs/1702.01113} {arXiv:1702.01113 [hep-lat]}
  \BibitemShut {NoStop}%
\bibitem [{\citenamefont {Sievert}\ and\ \citenamefont
  {Noronha-Hostler}(2019)}]{Sievert:2019zjr}%
  \BibitemOpen
  \bibfield  {author} {\bibinfo {author} {\bibfnamefont {M.~D.}\ \bibnamefont
  {Sievert}}\ and\ \bibinfo {author} {\bibfnamefont {J.}~\bibnamefont
  {Noronha-Hostler}},\ }\href {\doibase 10.1103/PhysRevC.100.024904} {\bibfield
   {journal} {\bibinfo  {journal} {Phys. Rev. C}\ }\textbf {\bibinfo {volume}
  {100}},\ \bibinfo {pages} {024904} (\bibinfo {year} {2019})},\ \Eprint
  {http://arxiv.org/abs/1901.01319} {arXiv:1901.01319 [nucl-th]} \BibitemShut
  {NoStop}%
\bibitem [{\citenamefont {Plumberg}\ \emph {et~al.}(2021)\citenamefont
  {Plumberg}, \citenamefont {Almaalol}, \citenamefont {Dore}, \citenamefont
  {Noronha},\ and\ \citenamefont {Noronha-Hostler}}]{Plumberg:2021bme}%
  \BibitemOpen
  \bibfield  {author} {\bibinfo {author} {\bibfnamefont {C.}~\bibnamefont
  {Plumberg}}, \bibinfo {author} {\bibfnamefont {D.}~\bibnamefont {Almaalol}},
  \bibinfo {author} {\bibfnamefont {T.}~\bibnamefont {Dore}}, \bibinfo {author}
  {\bibfnamefont {J.}~\bibnamefont {Noronha}}, \ and\ \bibinfo {author}
  {\bibfnamefont {J.}~\bibnamefont {Noronha-Hostler}},\ }\href@noop {} {\
  (\bibinfo {year} {2021})},\ \Eprint {http://arxiv.org/abs/2103.15889}
  {arXiv:2103.15889 [nucl-th]} \BibitemShut {NoStop}%
\bibitem [{\citenamefont {Cheng}\ and\ \citenamefont
  {Shen}(2021)}]{Cheng:2021tnq}%
  \BibitemOpen
  \bibfield  {author} {\bibinfo {author} {\bibfnamefont {C.}~\bibnamefont
  {Cheng}}\ and\ \bibinfo {author} {\bibfnamefont {C.}~\bibnamefont {Shen}},\
  }\href@noop {} {\  (\bibinfo {year} {2021})},\ \Eprint
  {http://arxiv.org/abs/2103.09848} {arXiv:2103.09848 [nucl-th]} \BibitemShut
  {NoStop}%
\bibitem [{\citenamefont {Noronha-Hostler}\ \emph
  {et~al.}(2016{\natexlab{b}})\citenamefont {Noronha-Hostler}, \citenamefont
  {Yan}, \citenamefont {Gardim},\ and\ \citenamefont
  {Ollitrault}}]{Noronha-Hostler:2015dbi}%
  \BibitemOpen
  \bibfield  {author} {\bibinfo {author} {\bibfnamefont {J.}~\bibnamefont
  {Noronha-Hostler}}, \bibinfo {author} {\bibfnamefont {L.}~\bibnamefont
  {Yan}}, \bibinfo {author} {\bibfnamefont {F.~G.}\ \bibnamefont {Gardim}}, \
  and\ \bibinfo {author} {\bibfnamefont {J.-Y.}\ \bibnamefont {Ollitrault}},\
  }\href {\doibase 10.1103/PhysRevC.93.014909} {\bibfield  {journal} {\bibinfo
  {journal} {Phys. Rev. C}\ }\textbf {\bibinfo {volume} {93}},\ \bibinfo
  {pages} {014909} (\bibinfo {year} {2016}{\natexlab{b}})},\ \Eprint
  {http://arxiv.org/abs/1511.03896} {arXiv:1511.03896 [nucl-th]} \BibitemShut
  {NoStop}%
\bibitem [{\citenamefont {Adamczyk}\ \emph
  {et~al.}(2015{\natexlab{b}})\citenamefont {Adamczyk} \emph
  {et~al.}}]{STAR:2015mki}%
  \BibitemOpen
  \bibfield  {author} {\bibinfo {author} {\bibfnamefont {L.}~\bibnamefont
  {Adamczyk}} \emph {et~al.} (\bibinfo {collaboration} {STAR}),\ }\href
  {\doibase 10.1103/PhysRevLett.115.222301} {\bibfield  {journal} {\bibinfo
  {journal} {Phys. Rev. Lett.}\ }\textbf {\bibinfo {volume} {115}},\ \bibinfo
  {pages} {222301} (\bibinfo {year} {2015}{\natexlab{b}})},\ \Eprint
  {http://arxiv.org/abs/1505.07812} {arXiv:1505.07812 [nucl-ex]} \BibitemShut
  {NoStop}%
\bibitem [{\citenamefont {Carzon}\ \emph {et~al.}(2020)\citenamefont {Carzon},
  \citenamefont {Rao}, \citenamefont {Luzum}, \citenamefont {Sievert},\ and\
  \citenamefont {Noronha-Hostler}}]{Carzon:2020xwp}%
  \BibitemOpen
  \bibfield  {author} {\bibinfo {author} {\bibfnamefont {P.}~\bibnamefont
  {Carzon}}, \bibinfo {author} {\bibfnamefont {S.}~\bibnamefont {Rao}},
  \bibinfo {author} {\bibfnamefont {M.}~\bibnamefont {Luzum}}, \bibinfo
  {author} {\bibfnamefont {M.}~\bibnamefont {Sievert}}, \ and\ \bibinfo
  {author} {\bibfnamefont {J.}~\bibnamefont {Noronha-Hostler}},\ }\href
  {\doibase 10.1103/PhysRevC.102.054905} {\bibfield  {journal} {\bibinfo
  {journal} {Phys. Rev. C}\ }\textbf {\bibinfo {volume} {102}},\ \bibinfo
  {pages} {054905} (\bibinfo {year} {2020})},\ \Eprint
  {http://arxiv.org/abs/2007.00780} {arXiv:2007.00780 [nucl-th]} \BibitemShut
  {NoStop}%
\bibitem [{\citenamefont {Aidala}\ \emph {et~al.}(2018)\citenamefont {Aidala}
  \emph {et~al.}}]{Aidala:2017ajz}%
  \BibitemOpen
  \bibfield  {author} {\bibinfo {author} {\bibfnamefont {C.}~\bibnamefont
  {Aidala}} \emph {et~al.} (\bibinfo {collaboration} {PHENIX}),\ }\href
  {\doibase 10.1103/PhysRevLett.120.062302} {\bibfield  {journal} {\bibinfo
  {journal} {Phys. Rev. Lett.}\ }\textbf {\bibinfo {volume} {120}},\ \bibinfo
  {pages} {062302} (\bibinfo {year} {2018})},\ \Eprint
  {http://arxiv.org/abs/1707.06108} {arXiv:1707.06108 [nucl-ex]} \BibitemShut
  {NoStop}%
\bibitem [{ATL(2021)}]{ATLAS:2021kty}%
  \BibitemOpen
  \href@noop {} {\  (\bibinfo {year} {2021})}\BibitemShut {NoStop}%
\bibitem [{\citenamefont {Jia}\ \emph {et~al.}(2016)\citenamefont {Jia},
  \citenamefont {Radhakrishnan},\ and\ \citenamefont {Zhou}}]{Jia:2015jga}%
  \BibitemOpen
  \bibfield  {author} {\bibinfo {author} {\bibfnamefont {J.}~\bibnamefont
  {Jia}}, \bibinfo {author} {\bibfnamefont {S.}~\bibnamefont {Radhakrishnan}},
  \ and\ \bibinfo {author} {\bibfnamefont {M.}~\bibnamefont {Zhou}},\ }\href
  {\doibase 10.1103/PhysRevC.93.044905} {\bibfield  {journal} {\bibinfo
  {journal} {Phys. Rev. C}\ }\textbf {\bibinfo {volume} {93}},\ \bibinfo
  {pages} {044905} (\bibinfo {year} {2016})},\ \Eprint
  {http://arxiv.org/abs/1506.03496} {arXiv:1506.03496 [nucl-th]} \BibitemShut
  {NoStop}%
\bibitem [{\citenamefont {Jia}\ \emph {et~al.}(2020)\citenamefont {Jia},
  \citenamefont {Zhang},\ and\ \citenamefont {Xu}}]{Jia:2020tvb}%
  \BibitemOpen
  \bibfield  {author} {\bibinfo {author} {\bibfnamefont {J.}~\bibnamefont
  {Jia}}, \bibinfo {author} {\bibfnamefont {C.}~\bibnamefont {Zhang}}, \ and\
  \bibinfo {author} {\bibfnamefont {J.}~\bibnamefont {Xu}},\ }\href {\doibase
  10.1103/PhysRevResearch.2.023319} {\bibfield  {journal} {\bibinfo  {journal}
  {Phys. Rev. Res.}\ }\textbf {\bibinfo {volume} {2}},\ \bibinfo {pages}
  {023319} (\bibinfo {year} {2020})},\ \Eprint
  {http://arxiv.org/abs/2001.08602} {arXiv:2001.08602 [nucl-th]} \BibitemShut
  {NoStop}%
\end{thebibliography}%

\end{document}